\pdfoutput=1
\documentclass[11pt,twoside,a4paper,cmspaper,final,collab]{cms-tdr}
\def\svnVersion{f604f65}\def\svnDate{2021/10/20}\def\cmsCernNoTag{CERN-EP-2021-131}\def\cmsCernDate{\today}\def\cmsMessage{Published in the Journal of High Energy Physics as \href{http://dx.doi.org/10.1007/JHEP11(2021)111}{\doi{10.1007/JHEP11(2021)111}.}}
\begin{document}\cmsNoteHeader{TOP-20-002}

\newlength\cmsTabSkip\setlength{\cmsTabSkip}{1ex}
\newcommand{\Lumimine}{\ensuremath{36\fbinv}}
\newcommand{\tW}{\ensuremath{\PQt\PW}\xspace}
\newcommand{\pp}{\ensuremath{\Pp\Pp}\xspace}
\newcommand{\wPlusJets}{\ensuremath{\PW\text{+jets}}\xspace}
\newcommand{\zPlusJets}{\ensuremath{\PZ\text{+jets}}\xspace}
\newcommand{\mtop}{\ensuremath{m_{\PQt}}\xspace}
\newcommand{\mtw}{\ensuremath{\smash{m_{\mathrm{T}}^{\PW}}\xspace}}
\newcommand{\wtb}{\ensuremath{\PW\PQt\PQb}\xspace}
\newcommand{\vtb}{\ensuremath{\PV_{\PQt\PQb}}\xspace}
\newcommand{\xsecannnlo}{\ensuremath{79.5\,^{+1.9}_{-1.8}\,\text{(scale)}\,^{+2.0}_{-1.4}\,\text{(PDF)}\unit{pb}}\xspace}
\newcommand{\xsecannlo}{\ensuremath{71.7\pm 1.8\,\text{(scale)}\pm3.4\,\text{(PDF)}\unit{pb}}\xspace}
\newcommand{\xsecmes}{\ensuremath{89\pm4\stat\pm12\syst\unit{pb}}\xspace}
\newcommand{\hdamp}{\ensuremath{h_\text{damp}}\xspace}
\newcommand{\VV}{\ensuremath{\PV\PV}\xspace}
 
\cmsNoteHeader{TOP-20-002} \title{Observation of \texorpdfstring{\tW}{tW} production in the single-lepton channel in \texorpdfstring{\pp collisions at $\sqrt{s} = 13\TeV$}{pp collisions at sqrt(s)=13 TeV}}

\date{\today}

\abstract{
A measurement of the cross section of the associated production of a single top quark and a \PW boson in final states with a muon or electron and jets in proton-proton collisions at $\sqrt{s}=13\TeV$ is presented. The data correspond to an integrated luminosity of \Lumimine collected with the CMS detector at the CERN LHC in 2016. A boosted decision tree is used to separate the \tW signal from the dominant \ttbar background, whilst the subleading \wPlusJets and multijet backgrounds are constrained using data-based estimates. This result is the first observation of the \tW process in final states containing a muon or electron and jets, with a significance exceeding 5 standard deviations. The cross section is determined to be \xsecmes, consistent with the standard model.
}

\hypersetup{pdfauthor={CMS Collaboration},%
pdftitle={Observation of tW production in the single-lepton channel in pp collisions at sqrt(s)=13 TeV},%
pdfsubject={CMS},%
pdfkeywords={CMS, tW, top quark, cross section}}

\maketitle 
\section{Introduction}
\label{sec:intro}

The observation of singly produced top quarks by the D0~\cite{PhysRevLett.103.092001} and CDF~\cite{PhysRevLett.103.092002} Collaborations opened a new era in the study of electroweak interactions of top quarks.
At hadron colliders, top quarks are produced predominantly via the strong interaction with an antiquark partner (\ttbar). 
Much less frequently, top quarks and antiquarks are produced singly by the electroweak interaction via the \wtb vertex between the \PW boson and the top and bottom quarks.
Three main processes contribute to electroweak single top quark production: 
the $t$ channel~\cite{ATLAS:2016qhd,CMS:2018lgn,ATLAS:2019hhu}, produced by quark scattering via the exchange of a virtual {\PW} boson;
the $s$ channel~\cite{ATLAS:2015jmq,CMS:2016xoq}, produced by quark-antiquark annihilation to an off-shell {\PW} boson;
and the associated production of a single top quark with a {\PW} boson (\tW), produced either via the exchange of a top quark or by an intermediate off-shell {\PQb} quark. 

All three single top quark processes are sensitive to the Cabibbo--Kobayashi--Maskawa matrix element \vtb, and their study provides a direct probe of its value.
Any significant deviation from the established value may be indicative of physics beyond the standard model (SM).
The \tW process is sensitive in particular to the \wtb vertex, whilst the $t$- and $s$-channel processes contain contributions from additional four-fermion operators.
By studying all three single top quark channels it should, therefore, be possible to disentangle the new physics effects, if any such deviations are observed~\cite{Tait:2000sh,Cao:2007ea}. 

Whilst the Fermilab Tevatron experiments successfully observed the $t$- and $s$-channel processes~\cite{PhysRevLett.103.092001,PhysRevLett.103.092002}, the \tW production cross section was too small to be accessible.
At the CERN LHC, the \tW process has the second-largest cross section among the single top quark channels after the $t$ channel, making detailed studies of the \tW process possible.
Evidence of the \tW process was first reported by the ATLAS and CMS experiments at the LHC using data at $\sqrt{s}=7\TeV$~\cite{Aad:2012xca,Chatrchyan:2012zca}, followed by the observation at $\sqrt{s}=8\TeV$~\cite{Chatrchyan:2014tua,Aad:2015eto}.
Precise cross section and differential measurements have since been carried out using data at $\sqrt{s}=13\TeV$ by both collaborations~\cite{Aaboud:2016lpj,Aaboud:2017qyi,Sirunyan:2018lcp}.

\begin{figure}[h]
  \centering
    \includegraphics[width=0.27\textwidth]{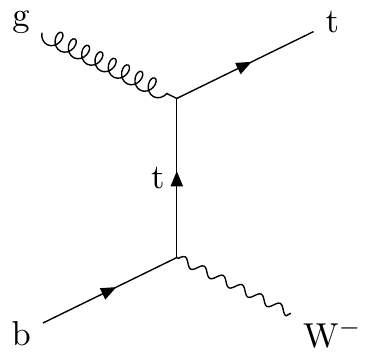}
    \includegraphics[width=0.27\textwidth]{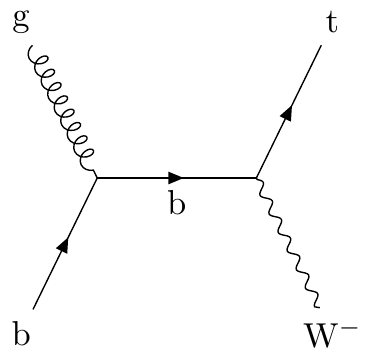}
\caption{\label{fig:fdlo}Leading-order Feynman diagrams for single top quark production in the \tW channel. 
      Charge conjugate states are implied.} 
\end{figure}

The leading-order (LO) Feynman diagrams for the \tW process are shown in Fig.~\ref{fig:fdlo}.
The production cross section in proton-proton (\pp) collisions at $\sqrt{s}=13\TeV$, assuming a top quark mass \mtop of 172.5\GeV, has been computed to be \xsecannlo at approximate next-to-next-to-LO (NNLO)~\cite{Kidonakis:2015nna}, and \xsecannnlo at approximate next-to-NNLO (aN$^{3}$LO)~\cite{Kidonakis_2021}.
The first uncertainties are due to scale variations in the calculation, and the second correspond to the choice of parton distribution functions (PDFs). 

\begin{figure}[h]
  \centering
    \includegraphics[width=0.3\textwidth]{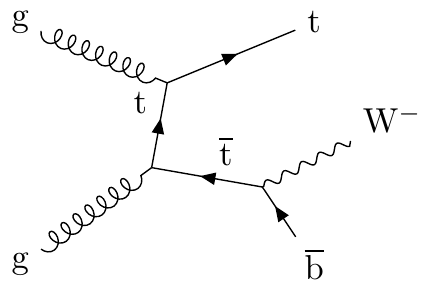}
    \includegraphics[width=0.3\textwidth]{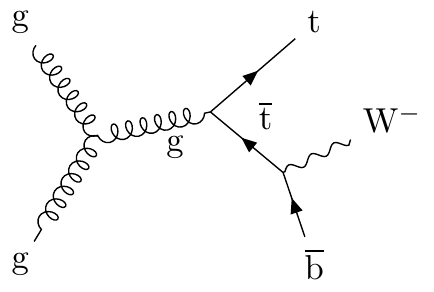}
    \includegraphics[width=0.3\textwidth]{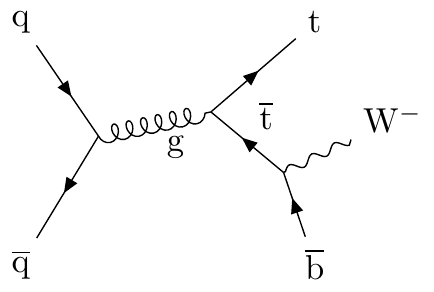}
\caption{\label{fig:fdnlo}Feynman diagrams for \tW single top quark production at next-to-leading order that are removed from the signal definition in the DR scheme.
      Charge conjugate states are implied.}
\end{figure}

The \tW process is of special interest because of its interference at next-to-LO (NLO) with \ttbar production~\cite{PhysRevD.59.075001,White:2009yt,Frixione:2008yi}.
Whilst the two processes are distinct at LO, they share  a subset of Feynman diagrams at NLO, examples of which can be seen in Fig.~\ref{fig:fdnlo}.
This leads to conceptual and practical problems with signal definition, the understanding and measurement of which can provide insight into how such types of  interference predicted in various new physics models might manifest.
Two schemes have been proposed to describe the \tW signal: ``diagram  removal'' (DR)~\cite{Frixione:2008yi}, where all NLO diagrams that are doubly resonant, such as those in Fig.~\ref{fig:fdnlo}, are excluded from the signal definition; and ``diagram subtraction'' (DS)~\cite{Frixione:2008yi,Tait:1999cf}, in which the differential cross section is modified with a gauge-invariant subtraction term that locally cancels the contribution of the \ttbar diagrams.
The DR scheme is used to define the \tW signal in this analysis.

In the SM, top quarks decay almost exclusively to a \PW boson and a \PQb quark.
Consequently, the \tW process results in a signature containing two \PW bosons and one \PQb quark.

To date, all \tW studies carried out on data collected by the CMS detector have been performed using the final states in which both \PW bosons decay leptonically.
In comparison to this well-established final state, the single-lepton final state--in which one \PW boson decays leptonically and the other hadronically--has seen little study; to date, only one measurement has been presented by the ATLAS Collaboration using data at $\sqrt{s}=8\TeV$~\cite{atlascollaboration2020measurement}.
Whilst the single-lepton channel offers the advantages of larger branching fractions and the possibility of a fully reconstructable top quark system, it suffers from larger and more numerous backgrounds.

This paper reports the first measurement from the CMS Collaboration of the \tW process in the single-lepton final state. 
Single-lepton events are selected from \pp collisions at $\sqrt{s}=13\TeV$ corresponding to an integrated luminosity of \Lumimine.
A boosted decision tree (BDT) is used to separate the \tW signal from the dominant \ttbar background.
The subdominant \wPlusJets events and events comprised of jets produced through the strong interaction, referred to as quantum chromodynamic (QCD) multijet events, are constrained using data-based estimates.
The \tW production cross section is extracted using a binned likelihood fit carried out on the BDT discriminant distributions for both channels and three jet multiplicity regions simultaneously.
Tabulated results are provided in HEPData~\cite{hepDataEntry}.

\section{The CMS detector}
\label{EXsetup}

The central feature of the CMS apparatus is a superconducting solenoid of 6\unit{m} internal diameter, providing a magnetic field of 3.8\unit{T}. Within the solenoid volume are a silicon pixel and strip tracker, a lead tungstate crystal electromagnetic calorimeter (ECAL), and a brass and scintillator hadron calorimeter (HCAL), each composed of a barrel and two endcap sections. Forward calorimeters extend the pseudorapidity ($\eta$) coverage provided by the barrel and endcap detectors. Muons are detected in gas-ionization chambers embedded in the steel flux-return yoke outside the solenoid. 

The candidate vertex with the largest value of summed physics-object $\pt^2$ (where \pt is the transverse momentum) is taken to be the primary \pp interaction vertex. The physics objects are the jets, clustered using the jet finding algorithm~\cite{Cacciari:2008gp,Cacciari:2011ma} with the tracks assigned to candidate vertices as inputs, and the associated missing transverse momentum, taken as the negative vector \pt sum of those jets.

The particle-flow algorithm~\cite{CMS-PRF-14-001} aims to reconstruct and identify each individual particle in an event, with an optimized combination of information from the various elements of the CMS detector. The energy of photons is obtained from the ECAL measurement. The energy of electrons is determined from a combination of the electron momentum at the primary interaction vertex as determined by the tracker, the energy of the corresponding ECAL cluster, and the energy sum of all bremsstrahlung photons spatially compatible with originating from the electron track. The energy of muons is obtained from the curvature of the corresponding track~\cite{muonRecoPaper}. The energy of charged hadrons is determined from a combination of their momentum measured in the tracker and the matching ECAL and HCAL energy deposits, corrected for the response function of the calorimeters to hadronic showers. Finally, the energy of neutral hadrons is obtained from the corresponding corrected ECAL and HCAL energies. 

The missing transverse momentum vector \ptvecmiss is computed as the negative vector \pt sum of all the particle-flow candidates in an event, and its magnitude is denoted as \ptmiss~\cite{Sirunyan:2019kia}. The \ptvecmiss is modified to account for corrections to the energy scale of the reconstructed jets in the event.

A more detailed description of the CMS detector, together with a definition of the coordinate system used and the relevant kinematic variables, can be found in Ref.~\cite{Chatrchyan:2008zzk}. 

\section{Data and simulated samples}

The measurement uses data collected with the CMS detector during \pp collisions in 2016 at $\sqrt{s}=13\TeV$, corresponding to an integrated luminosity of \Lumimine~\cite{lumiPaper}.

Events simulated using the Monte Carlo (MC) method are used throughout the analysis. 
Signal \tW events are simulated using the \POWHEG v1~\cite{Re:2010bp} generator interfaced with \PYTHIA 8.205~\cite{Sjostrand:2014zea} for showering using the CUETP8M1 tune~\cite{Khachatryan:2015pea}.
Fully hadronic decays are excluded from the simulation, and separate samples are created for top quark and antiquark events.
The \tW process signal is defined using the DR scheme. 
Events for the \ttbar background are simulated using \POWHEG v2~\cite{Alioli:2010xd} interfaced with \PYTHIA 8.205 using the CUETPM2T4 tune~\cite{Skands:2014pea}.
The second-leading background, \wPlusJets, is simulated using \MGvATNLO 2.2.2~\cite{Alwall:2014hca}.
The matrix element (ME) calculations are matched to parton shower (PS) using the FxFx~\cite{Frederix:2012ps} algorithm.
Single top quark backgrounds from the $t$ and $s$ channel--together referred to as the single \PQt background throughout this paper--are generated using \POWHEG v2 interfaced with \PYTHIA 8.205 with the CUETP8M1 tune, including spin correlations~\cite{Artoisenet:2012st}.
QCD multijet events are simulated using \MGvATNLO interfaced with \PYTHIA 8.205 using the MLM matching~\cite{Alwall:2007fs}.
The $\PW\PW$, $\PW\PZ$ and $\PZ\PZ$ diboson backgrounds--collectively referred to as the \VV background--are simulated using \PYTHIA 8.205 with the CUETP8M1 tune.
All samples are generated at NLO in QCD with the exception of the \VV and QCD multijet processes, which are produced at LO.
Contributions from other processes are found to be negligible.

For all samples, the proton structure is described using the NNPDF3.0~\cite{Ball:2012cx} PDF set, and \mtop is chosen to be 172.5\GeV.
Minimum bias \pp interactions generated using \PYTHIA 8.205 are overlayed on all simulated events to account for additional interactions occuring per bunch crossing that do not originate from the primary vertex of interest (pileup).
The detector response is simulated using the \GEANTfour package~\cite{AGOSTINELLI2003250,1610988}.

All simulated events are processed using the same software chain as for collision data, reweighted to account for the observed distribution in pileup, and normalized to the predicted cross section of the process.

 \section{Event selection}
\label{Evtselect}

Events of interest are selected using a two-tiered trigger system~\cite{Sirunyan:2020zal,Khachatryan:2016bia}.
To be considered for the analysis, events must pass high-level triggers that select a single lepton with \pt of at least 24 (27)\GeV  for muons (electrons).
Additional offline selections are made such that each event contains exactly one muon with $\pt>26\GeV$ and $\abs{\eta} < 2.1$ or one electron with $\pt > 30\GeV$ and $\abs{\eta} < 1.48$.
The forward $\eta$ range is excluded from the electron selection because background processes dominate in this region.
These leptons must pass identification and isolation requirements~\cite{Khachatryan:2015hwa,muonRecoPaper}, and have originated from the well-reconstructed primary interaction vertex.
The isolation requirements are based on the ratio between the lepton \pt and the scalar sum of the \pt of charged hadrons and neutral particles within a cone of $\DR = \sqrt{\smash[b]{(\Delta \phi)^2+(\Delta \eta)^2}}=0.3$ of the lepton (corrected for pileup), where $\phi$ is the azimuthal angle in radians. 
Events that contain additional leptons with lower \pt requirements ($\pt > 10\GeV$ for muons and $\pt>20\GeV$ for electrons) and $\abs{\eta}<2.4$ are rejected.
Corrections are applied to the trigger and lepton efficiencies in simulation to match those observed in data. 

Further selections are made based on the jet topology of the event.
Particle-flow jets, reconstructed using the anti-\kt algorithm~\cite{Cacciari:2008gp} with a distance parameter $R=0.4$,
are selected if they have $\pt > 30\GeV$ and $\abs{\eta} < 2.4$.
Only jets that are $\DR > 0.4$ from the selected leptons are considered.
At least two and no more than four jets must be present in the event to be considered in the analysis.
The energy of the jets is corrected to take into account inefficiencies and anisotropies in the detectors and reconstruction stages~\cite{Khachatryan:2016kdb}.

Jets originating from the hadronization of a \PQb quark are identified ({\PQb}-tagged) using the combined secondary vertex v2 (CSVv2) algorithm~\cite{Sirunyan:2018bby}.
The candidate \PQb jets must pass the nominal jet selections, as well as a working point of the CSVv2 algorithm chosen to give a \PQb tagging efficiency of ${\approx}75\%$ for \PQb quark jets and a misidentification probability of 1\% for \PQu, \PQd, \PQs quark and gluon jets.
Exactly one jet that passes these criteria must be present in an event to be used in the analysis.
The \PQb tagging efficiencies and misidentification probability are corrected in simulation to match those observed in data.

No selection requirements are made on the \ptmiss of the event.
 \section{Analysis strategy}
\label{anaStrategy}

Events used in the final fit are classified into three distinct analysis regions, one signal region and two control regions.
Along with the requirements on leptons and \PQb tagging, an event must contain exactly three jets to be selected in the signal region (3j).

Two control regions are defined such that they are enhanced in the leading backgrounds of the analysis.
To keep the regions as kinematically similar to the signal as possible, the selection requirements applied to these regions are identical to those of the signal region, 
with the exception of the number of selected jets.
The first such region contains events with exactly two jets (2j), and is enhanced in the \wPlusJets and QCD multijet backgrounds.
The second contains events with exactly four jets (4j), and is enhanced in \ttbar background.

Normalized distributions (templates) and normalization estimates for all processes are taken directly from simulation, with the exception of the \wPlusJets and QCD multijet backgrounds.
In the case of the \wPlusJets background, templates are taken from simulation but with the normalization corrected using data to account for the observed mismodelling of jet composition in simulation with respect to data.
For the QCD multijet background, mismodelling in both genuine leptons produced in hadron decays, and photon conversions and other objects incorrectly identified as leptons--collectively referred to as nonprompt leptons--precipitates the need to extract both templates and normalization estimates from data directly.

By far the largest contribution to the QCD multijet background is found to be when a jet contains a nonprompt lepton and therefore passes the signal selection requirements.
In order to model this background, a sample enriched in these nonprompt leptons is defined.
By inverting the isolation requirement on the selected lepton, a sample that is dominated by the QCD multijet background can be created that is as kinematically similar to the desired analysis regions as possible whilst remaining statistically independent.
Templates to be used in the final fit of the analysis regions are extracted from these events.
A small contribution of \ttbar events is found in this sample, and their contribution--estimated from simulation--is subtracted before use.

The normalizations of both the QCD multijet and \wPlusJets backgrounds are then estimated together using a binned likelihood fit on a distribution that has good separating power between the two processes.
The chosen distribution is the transverse mass \mtw of the reconstructed leptonically decaying \PW boson candidate, defined as
\begin{equation}
\mtw = \sqrt{2  \ptmiss  \pt^{\ell} \left(1-\cos[\phi^{\ptvecmiss} - \phi^\ell]\right)},
\end{equation}
where $\pt^{\ell}$ is the lepton \pt, and $\phi^{\ptvecmiss}$ and $\phi^\ell$ are the azimuthal angles of the \ptvecmiss and lepton, respectively.
In events with a real \PW boson, such as the \wPlusJets background, this distribution peaks at the \PW boson mass, whereas backgrounds with no real \PW boson, such as QCD multijet, exhibit a falling distribution that peaks at zero.
To avoid potential bias, the fit is carried out in a sample that is enhanced in \wPlusJets and QCD multijet events but statistically independent from the analysis regions, namely on a sample with exactly two jets, neither of which pass \PQb tagging requirements.
All other backgrounds are fixed to the values obtained from simulation.
Correction factors for both the \wPlusJets and QCD multijet processes are calculated by comparing the results of the fit with initial yield estimates taken directly from simulation.
These correction factors are then applied to the expected yields from simulation in each analysis region to estimate the normalization of the two backgrounds.

The uncertainty in extrapolating the correction factors to the analysis regions is assessed by performing the \mtw fit to the analysis regions (rather than the no-{\PQb}-tag sample), and treating the difference as the uncertainty.
Both this and the uncertainty from the fit are included in the normalization uncertainty of the \wPlusJets and QCD multijet processes in the final fit.

\begin{figure}[htb!]
  \centering
    \includegraphics[width=0.49\textwidth]{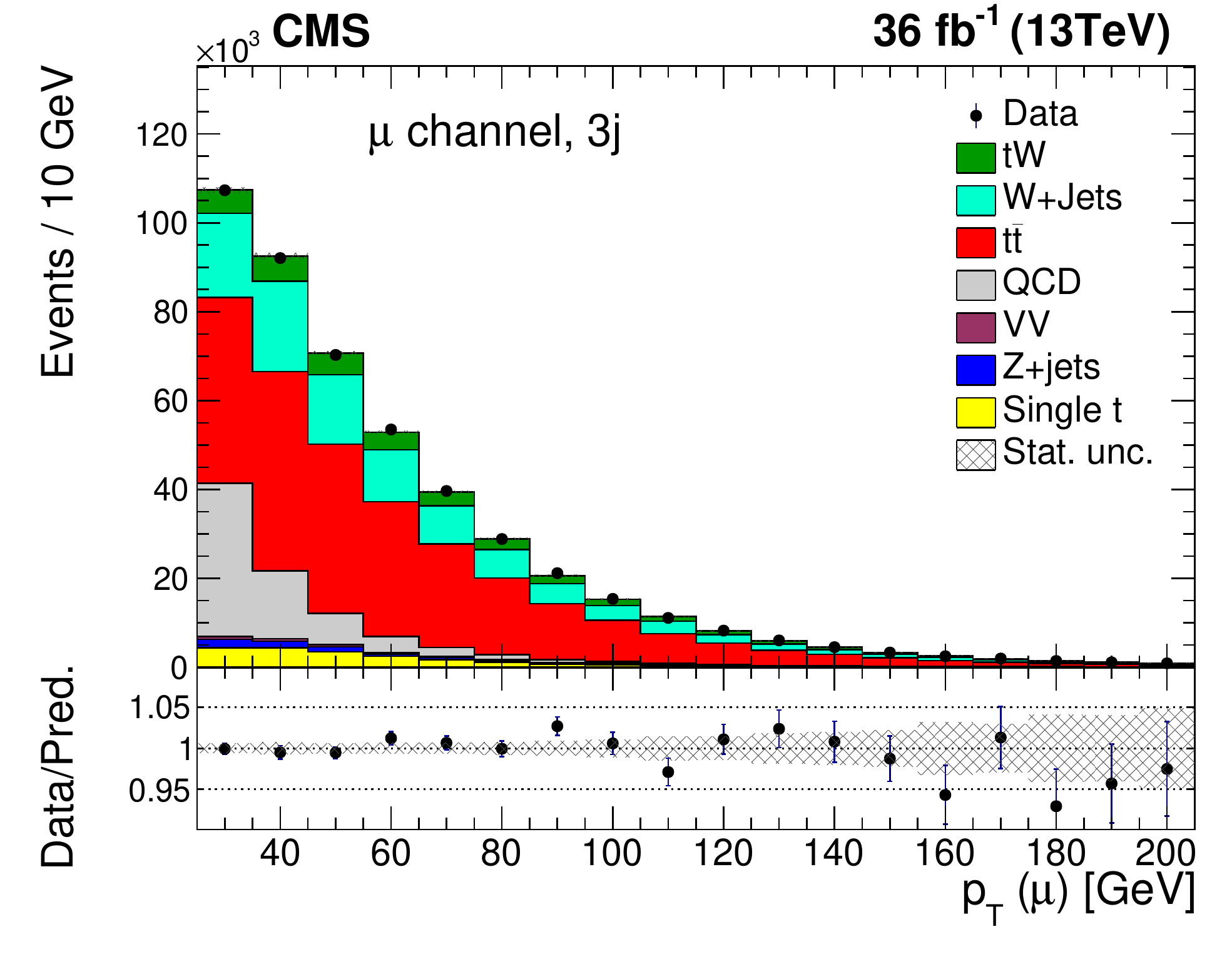}
    \includegraphics[width=0.49\textwidth]{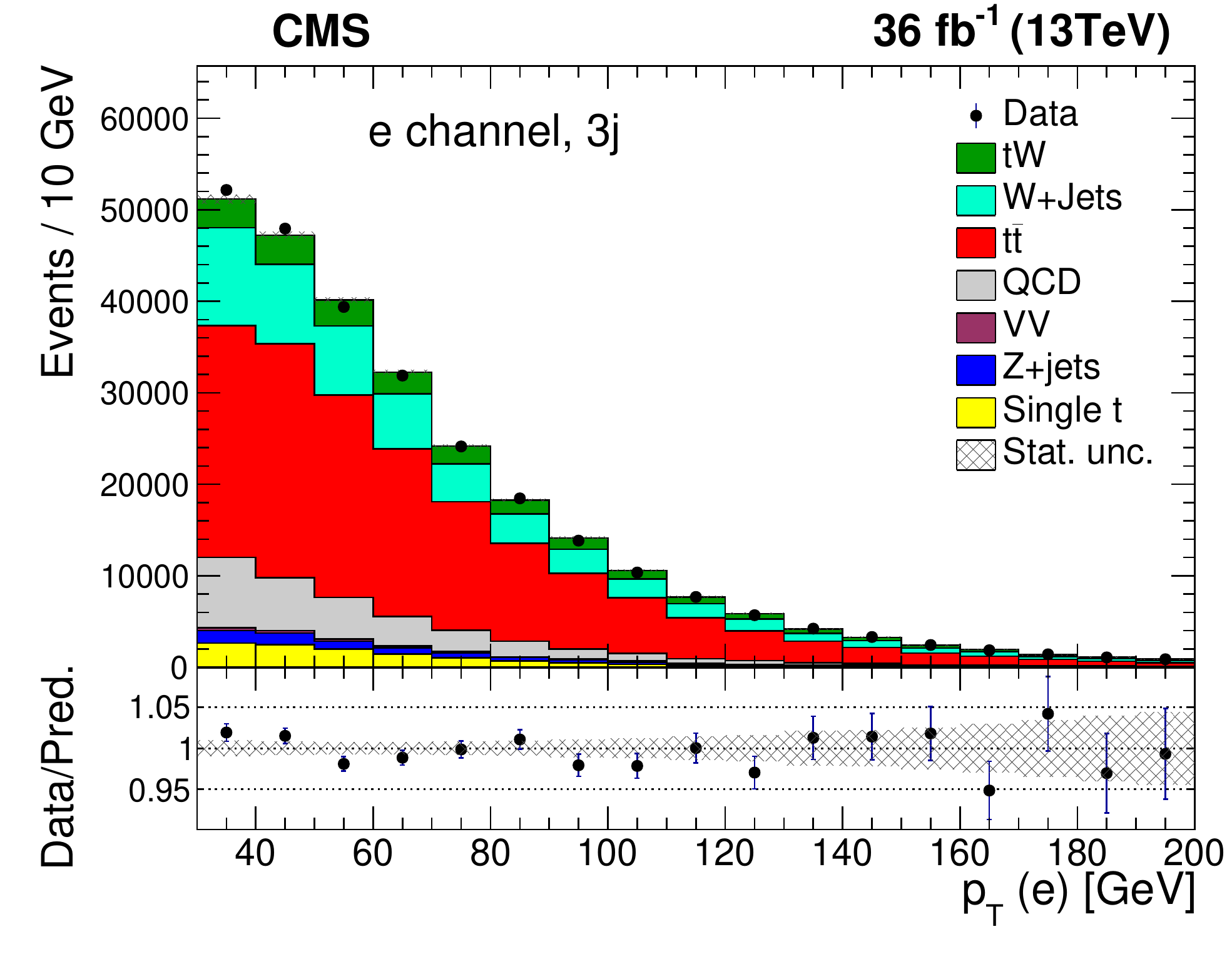}
    \caption{\label{fig:lepPt}The \pt of the selected muon (left) and electron (right) in the signal region of their respective channels.
      The signal and backgrounds have been scaled with the results of the final fit.
      The lower panel shows the ratio of observed data to the prediction for signal and background.
      In both panels the hatched regions show the statistical uncertainty from the limited size of the simulated samples for each bin.
      }
\end{figure}

Table~\ref{tab:eventYields} shows the event yields per process for each analysis region for the muon and electron channels. 
Figure~\ref{fig:lepPt} shows the \pt of the selected lepton in the signal region for the muon and electron channels, scaled to the result of the final fit.

\begin{table*}[ht!]
  \topcaption{\label{tab:eventYields}The total number of events passing the event selection in each analysis region and their associated statistical uncertainties.
    The event yields are given for the \tW signal and all major backgrounds for both the muon (upper) and electron (lower) channels.
    These values are provided for reference using simulation and scaled to the SM cross sections, with the exception of the QCD multijet background, which is taken from a data-based method, and the \wPlusJets background, which uses the SM cross section corrected using a data-based method. 
    A more precise estimation is obtained from the final fit, as described in the text.    
    The single \PQt background is comprised of the $t$- and $s$-channel single top quark processes.\\
  }
\centering
    \begin{tabular}{lc         
        r @{${}\pm{}$} l 
r @{${}\pm{}$} l 
r @{${}\pm{}$} l 
}
      \multirow{2}{*}{Sample} & & \multicolumn{6}{c}{Muon channel} \\
      & & \multicolumn{2}{c}{3j} & \multicolumn{2}{c}{2j} & \multicolumn{2}{c}{4j} \\
      \hline
      \tW           & & 26083 & 62 & 29814 & 66 & 10612 & 40 \\
      \ttbar        & & 274100 & 360 & 198120 & 300 & 186200 & 300 \\
      \wPlusJets    & & 79500 & 1200 & 319800 & 3200 & 18000 & 480 \\
      QCD multijet  & & 66830 & 360 & 277610 & 940 & 7700 & 110 \\
      Single \PQt      & & 15786 & 55 & 55250 & 100 & 4124 & 28 \\
      \zPlusJets    & & 7290 & 500 & 26950 & 960 & 2080 & 240 \\
      \VV            & & 2860 & 160 & 7480 & 250 & 754 & 83 \\
[\cmsTabSkip]
      Total prediction & & 472500 & 2700 & 915000 & 5800  & 229400 & 1300  \\
Data && \multicolumn{1}{l}{472540} && \multicolumn{1}{l}{923880} && \multicolumn{1}{l}{223720} & \\

      \multicolumn{8}{c}{} \\

      \multirow{2}{*}{Sample} & & \multicolumn{6}{c}{Electron channel}\\
& & \multicolumn{2}{c}{3j} & \multicolumn{2}{c}{2j} & \multicolumn{2}{c}{4j} \\
      \hline
\tW           & & 15726 & 35 & 17479 & 36 & 6596 & 23 \\
\ttbar        & & 156050 & 200 & 109980 & 160 & 108410 & 160 \\
      \wPlusJets    & & 50230 & 670 & 192400 & 1800 & 12090 & 310 \\
      QCD multijet  & & 21120 & 410 & 87880 & 680 & 2370 & 79 \\
      Single \PQt      & & 8937 & 30 & 30335 & 54 & 2379 & 15 \\
      \zPlusJets    & & 6960 & 300 & 24170 & 590 & 1840 & 140 \\
      \VV            & & 1635 & 84 & 4050 & 130 & 463 & 44 \\
      [\cmsTabSkip]
      Total prediction & & 260700 & 1700 & 466300 & 3500 & 134000 & 780 \\
      Data && \multicolumn{1}{l}{270330} & & \multicolumn{1}{l}{462940} & & \multicolumn{1}{l}{136190} & \\
\end{tabular}
\end{table*}

After all selection requirements have been applied, \tW signal events are selected with an efficiency of about 5\%, and constitute 6\% of the expected events in the signal region.
In order to increase the sensitivity of the measurement, a multivariate analysis is used to distinguish this signal from the backgrounds.
For this analysis, a BDT is trained to identify signal \tW events from the dominant \ttbar background.
The implementation of the BDT is provided by the ``Toolkit for Multivariate Data Analysis''~\cite{Hocker:2007ht}, and uses the gradient boosting algorithm~\cite{Yang:2005nz}.
Although a considerable fraction of the selected events in the signal region comes from QCD multijet and \wPlusJets backgrounds, it was found that, given the relatively small number of available training events for these samples, including contributions from these backgrounds in the samples used to train the BDT did not improve the sensitivity of the result.

The input variables to the BDT are chosen based on their ability to separate the signal from the \ttbar background and the quality of their modelling in simulation.
The chosen variables exploit the only difference between a \tW and \ttbar event at LO, \ie the number of jets originating from the fragmentation of a \PQb quark.
For a \ttbar event to pass the selection criteria in the signal region, one jet must be misidentified or otherwise fail reconstruction.
The loss of this jet causes various kinematic distributions to differ significantly between the two processes, and is particularly noticeable when looking at combinations of reconstructed objects from the selected events.
For example, the two selected non-\PQb{}-tagged jets in the event should, for \tW signal events, originate from the hadronic decay of a \PW boson.
In a \ttbar event, however, it is possible that the two jets originate from separate decays.
This combinatoric uncertainty means that distributions containing combinations of these objects (angular separation (\DR), total invariant mass, etc.) differ from those of the signal.
In order to extract these distributions, candidates for the two intermediate \PW bosons in the \tW signal are reconstructed from the selected objects in each event; a leptonically decaying \PW boson candidate is reconstructed from the selected lepton and \ptmiss in the event, and a hadronically decaying \PW boson candidate is reconstructed from the two non-\PQb{}-tagged jets.

The BDT input variables, chosen to exploit a variety of these properties, are:
\begin{itemize}
\item mass of the hadronically decaying \PW boson candidate,
\item invariant mass of the \PQb{}-tagged jet and the sub-leading (in \pt) non-\PQb{}-tagged jet,
\item angular separation between the two non-\PQb{}-tagged jets,
\item angular separation between the reconstructed leptonic \PW boson candidate and leading (in \pt) non-\PQb{}-tagged jet,
\item \pt of the selected lepton,
\item energy of the two non-\PQb{}-tagged jets,
\item angular separation between the \PQb{}-tagged jet and the selected lepton,
\item transverse momentum of the system made of the three jets, lepton and \ptmiss.
\end{itemize}

One BDT is trained for each lepton flavour (electron and muon) in its respective signal region using a subset of the selected \tW and \ttbar events as the signal and background samples, respectively.
Although they are trained separately, the two BDTs share the same input variables.
The trained BDT is then applied to data and simulated samples in each analysis region for its respective lepton flavour, and the produced distributions are used as templates in a likelihood fit to measure the production cross section of the \tW process.
In the analysis regions where these variables may not be well defined, \eg the angular separation of the two non-\PQb{}-tagged jets in the 2j control region, a default value is assigned to the input variable before the discriminant is calculated.

\section{Systematic uncertainties}
\label{sys}

The sources of systematic uncertainty considered in the analysis are classified as either experimental or modelling uncertainties.
These systematic uncertainties are included in the signal extraction as nuisance parameters of the likelihood fit, as an effect on the normalization and/or shape of the input templates.
The experimental and modelling uncertainties impact on both shape and normalization, whilst the uncertainty of the luminosity measurement and background normalization uncertainties affect the normalization only.

Experimental uncertainties originate from corrections applied to the MC simulation in order to correctly describe data, and have a number of sources.
Uncertainties in the total inelastic cross section~\cite{Sirunyan:2018nqx} are propagated to the result by varying the pileup reweighting applied to simulated samples.
The lepton energy scale uncertainty also incorporates the impact of uncertainties in the identification, isolation, and reconstruction efficiencies of the selected leptons.
Lepton trigger efficiencies are calculated separately and included as an independent source of uncertainty in the result.

The momentum of the reconstructed jets is varied based on the applied jet energy corrections (the jet energy scale uncertainty~\cite{Khachatryan:2016kdb}), and the jet energy resolution~\cite{CMS-PAS-JME-16-003} of the detector.
The impact of these variations is propagated to the \ptmiss in the result.
An additional uncertainty on the \ptmiss is calculated by varying the unclustered energy in the detector that make up the \ptmiss within their respective energy resolutions.

The uncertainties associated with the measured \PQb tagging efficiency and misidentification rate~\cite{Sirunyan:2018bby} are included independently for each flavour of reconstructed jet.

The uncertainty in the measurement of the integrated luminosity collected during 2016 \pp collisions, 2.5\%~\cite{lumiPaper}, is propagated to the result.

In addition, an uncertainty in the production cross section for each of the background processes is included.
For the $t$-channel single top quark and \ttbar processes this uncertainty is taken from their respective recent CMS measurements~\cite{ttbarCrossSection,tChanCrossSection}.
For the \wPlusJets and QCD multijet backgrounds this uncertainty is taken from the data-based background estimation.
All other backgrounds are assigned an uncertainty of 50\%.
The normalization uncertainties are treated as correlated across all analysis regions, with the exception of the data-based backgrounds, which are assigned uncorrelated uncertainties in each analysis region.

The modelling uncertainties originate from the choices in the generator parameters made during event simulation.
These uncertainties are assessed by comparing the templates produced from the nominal samples with templates derived from alternate samples generated with variations in these parameters.
These parameters include ME scale variations in the \tW signal \POWHEG simulation~\cite{Catani:2003zt}. The strong coupling parameter \alpS, which controls the factorization and renormalization scales at parton shower level, is varied to produce samples that reflect the uncertainty in both the initial- and final-state radiation produced by the \tW signal and leading \ttbar background. 

The \hdamp parameter in \POWHEG, which controls the scale of parton shower~\cite{Skands:2014pea} matching with the ME~\cite{Sirunyan:2019dfx}, and therefore regulates the damping of real emission in NLO calculations, is varied in dedicated samples for the \ttbar background.
The effect of the underlying event on the \ttbar background is estimated by varying several parameters that together control the recoil part of the event.
The impact of the choice of colour reconnection model on the \ttbar background~\cite{Argyropoulos:2014zoa,Christiansen:2015yqa} is also assessed in the result.

The uncertainty in the proton PDFs is taken into account by reweighting simulated events using variations of the NNPDF3.0 set~\cite{Ball:2014uwa}.
The envelope of these varied weights is taken as the uncertainty in the likelihood fit.

In order to assess the impact of the choice of using the DR or DS scheme when simulating the \tW signal events, an alternate signal sample is generated using the DS scheme.
The templates that are produced using this alternate sample are treated as the morphed templates under the DR/DS nuisance parameter.

The systematic uncertainties are applied to all relevant processes, signal and backgrounds alike, in exactly the same manner.
Their associated nuisance parameters are treated as correlated between all analysis regions in which they are applicable. 
Where the sources differ due to the lepton flavour (i.e. trigger efficiencies, lepton scale uncertainties), the three regions of each lepton flavour are correlated with each other, but uncorrelated from the regions of opposite flavour. 
The data-based background uncertainties are uncorrelated across all regions.

For the case of nuisance parameters that change the shape of the input templates, the morphed templates are smoothed with a polynomial fit in order to avoid unrealistic constraints originating from statistical fluctuations.
The contribution of each systematic source to the total uncertainty of the result is displayed in Table~\ref{tab:combinedNuisanceParameterContribs}.

\begin{table*}
\centering
\topcaption{\label{tab:combinedNuisanceParameterContribs} Relative uncertainty in the measured cross section from each source of systematic uncertainty for the combination of the muon and electron channels.
The table is divided into experimental, normalization, and modelling uncertainties. 
Uncertainties arising from the limited size of the simulated samples are included in the statistical uncertainty.\\
}
\begin{tabular}{lc}
Source & Relative uncertainty (\%) \\
\hline
\textit{Experimental} & \\
[\cmsTabSkip]
Jet energy scale & 6 \\ {\PQb} tagging efficiency & 4 \\Luminosity & 3 \\Lepton energy scale & 2 \\Trigger efficiency & 1 \\Jet energy resolution & 1 \\ {\PQb} tagging misidentification rate & $<$1 \\Unclustered energy & $<$1 \\Pileup & $<$1 \\[\cmsTabSkip]
\textit{Normalization} & \\
[\cmsTabSkip]
QCD multijet normalization & 7 \\ \wPlusJets normalization & 6 \\ \zPlusJets normalization & 3 \\Single \PQt normalization & 1 \\ \ttbar normalization & 1 \\\VV normalization & $<$1 \\[\cmsTabSkip]
\textit{Modelling} & \\
[\cmsTabSkip]
\hdamp & 4 \\Diagram removal/diagram subtraction & 3 \\Underlying event tune & 3 \\Colour reconnection model & 1 \\Parton distribution function & 1 \\Matrix element/parton shower matching & 1 \\Final-state radiation & $<$1 \\Initial-state radiation & $<$1 \\[\cmsTabSkip]
Total systematic uncertainty & 14 \\
Statistical uncertainty & 5 \\[\cmsTabSkip]
Total uncertainty & 15 \\
\end{tabular}
\end{table*}

\section{Results}
\label{results}

A binned likelihood fit is performed on the BDT discriminants in order to extract the \tW production cross section.
All regions in the muon and electron channels are fit simultaneously to produce the result, with systematic uncertainties included as nuisance parameters in the fit.

The likelihood used in the statistical analysis, $\mathcal{L}(\sigma,\vec{\theta})$, is a function of the measured signal cross section $\sigma$, and a set of nuisance parameters $\vec{\theta}$ that parameterise the systematic uncertainties as nuisance parameters associated with log-normal priors.
The number of events in each bin of the input templates is assumed to be described by a Poisson distribution, and is a function of the number of predicted background events, $\mu$, and $\vec{\theta}$.
The best value for $\mu$ is then found by maximising the likelihood with respect to all of its parameters.
The impact of each source of systematic uncertainty is assessed by performing the fit with the remaining nuisance parameters held constant.

The measured \tW production cross section is \xsecmes.
The total observed uncertainty on the measurement is 15\%, compared to an expected uncertainty of 17\%.
This result is compatible with both the SM predictions for the process of \xsecannlo at NNLO in QCD~\cite{Kidonakis:2015nna}, and \xsecannnlo at aN$^{3}$LO~\cite{Kidonakis_2021}.
This corresponds to an excess of signal over the background-only hypothesis that exceeds 5 standard deviations, and is therefore the first observation of the \tW channel in the single-lepton final state.

\begin{figure}[htbp!]
 \centering
 \includegraphics[width=0.49\textwidth]{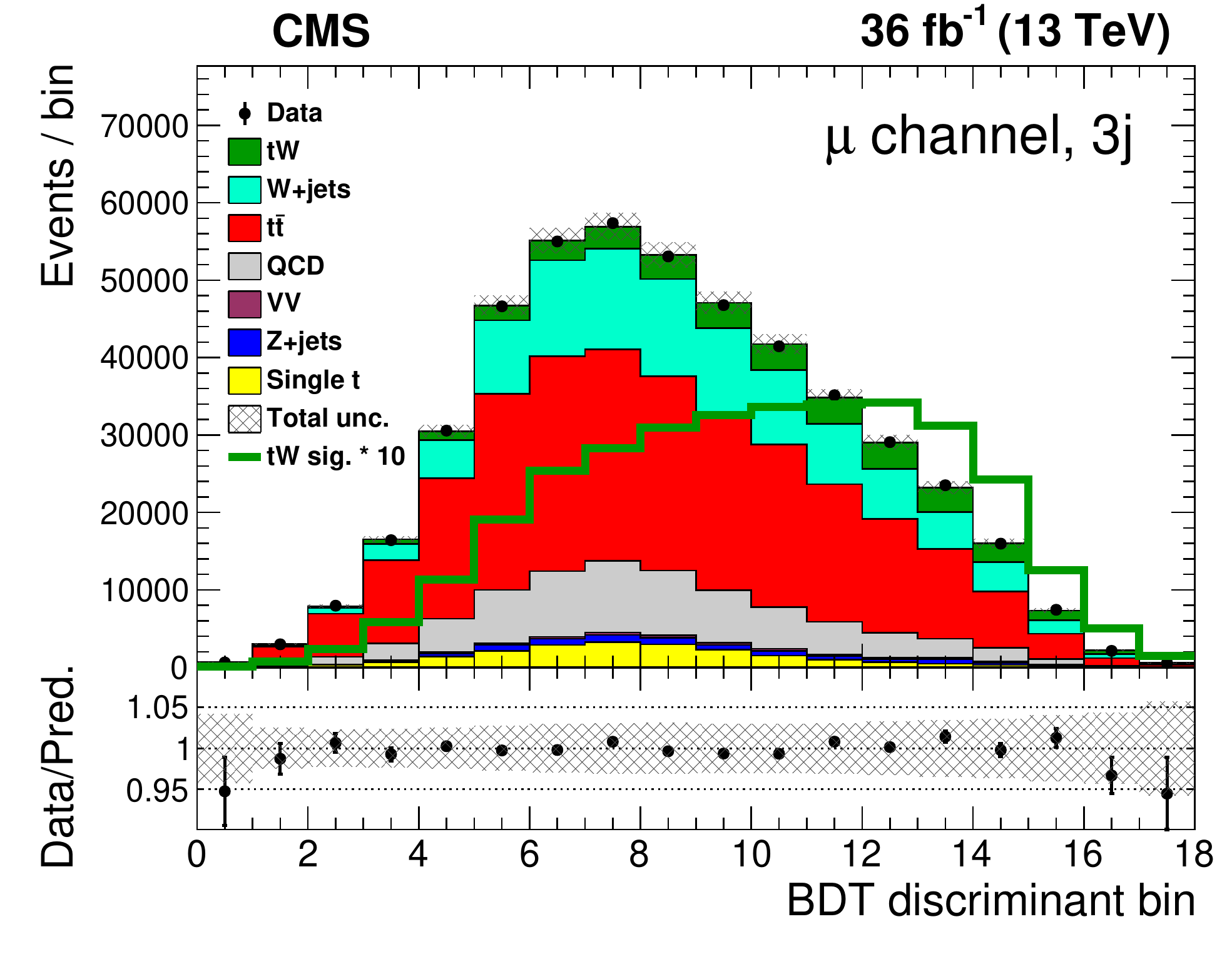}
 \includegraphics[width=0.49\textwidth]{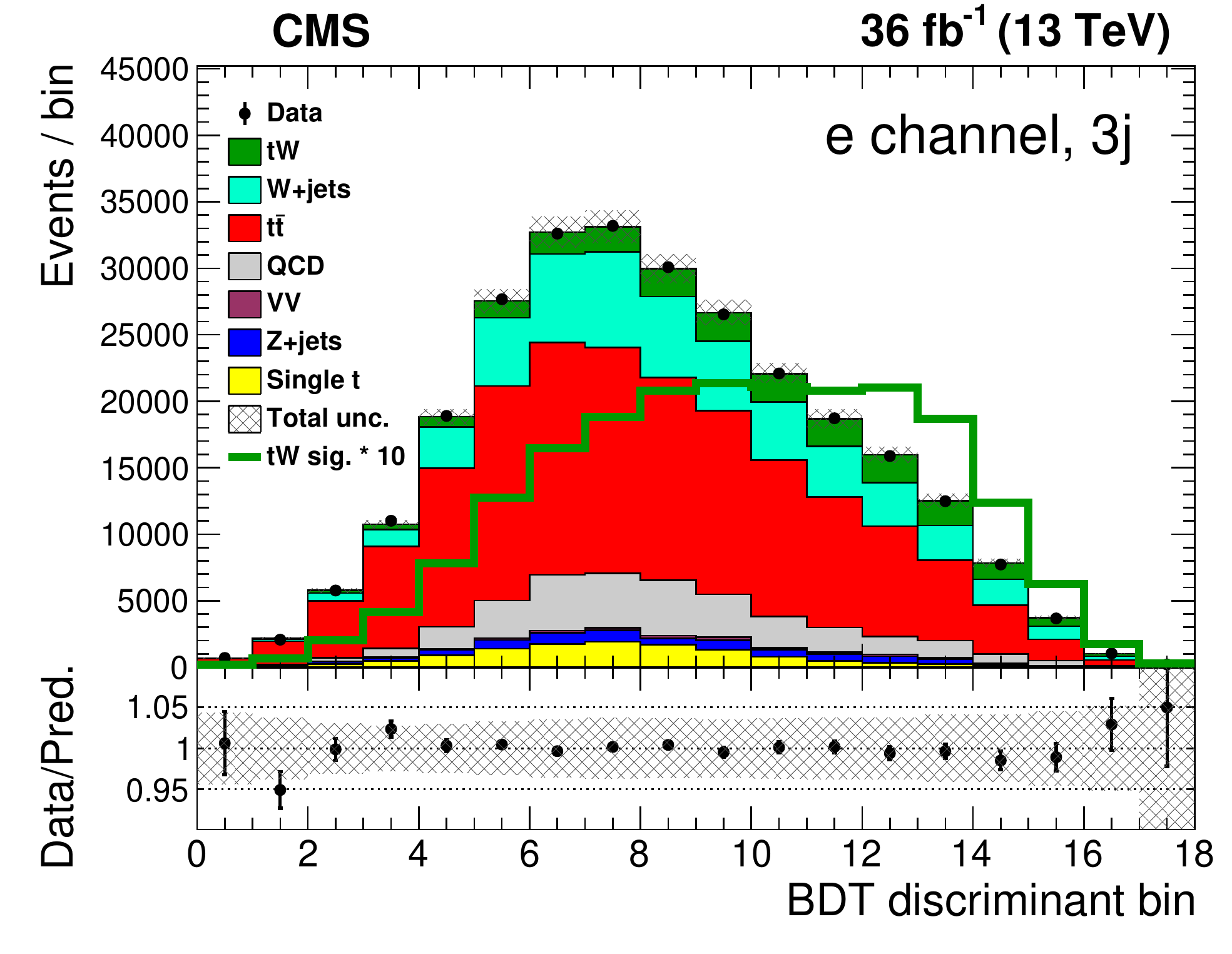} \\
 \includegraphics[width=0.49\textwidth]{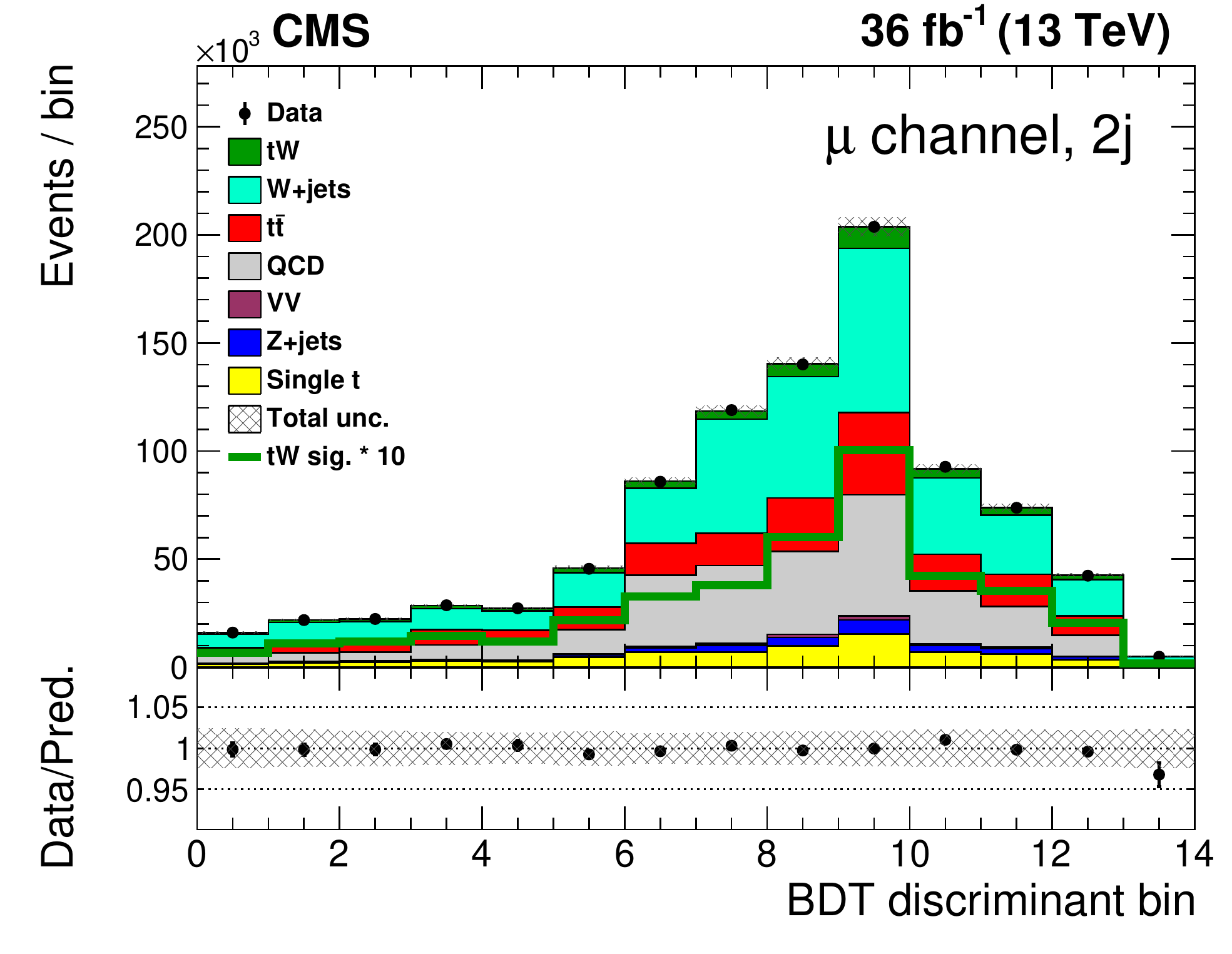}
 \includegraphics[width=0.49\textwidth]{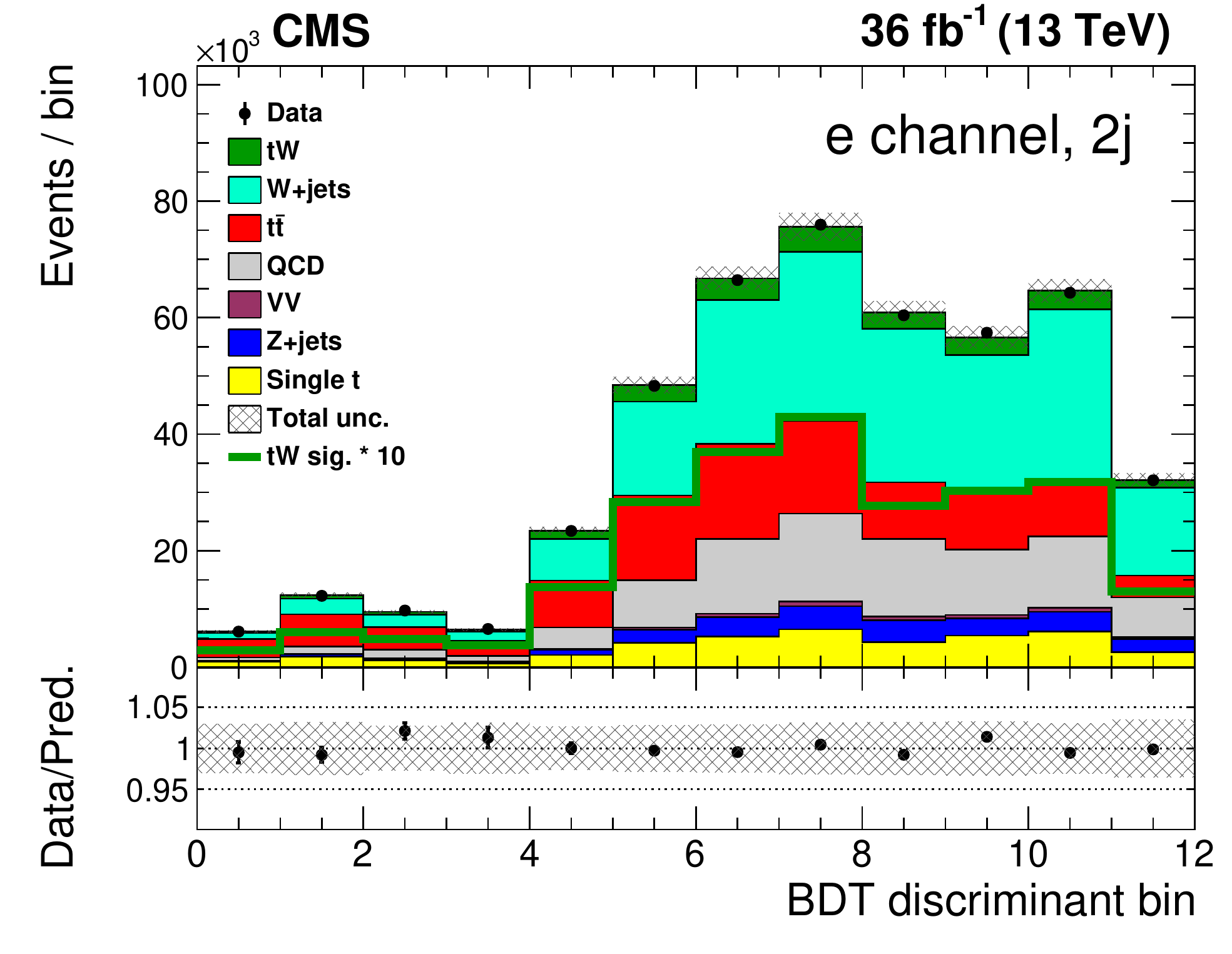} \\
 \includegraphics[width=0.49\textwidth]{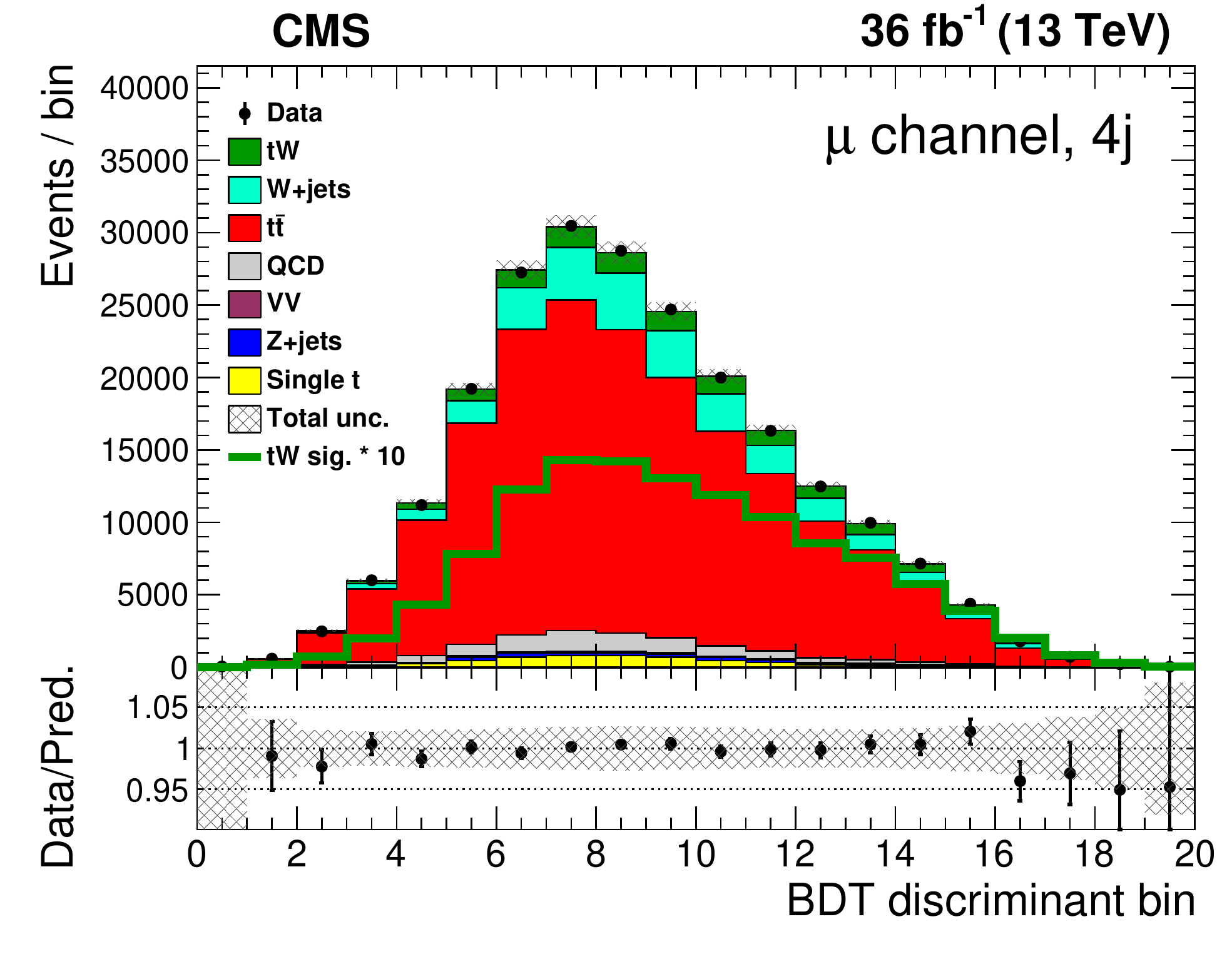}
 \includegraphics[width=0.49\textwidth]{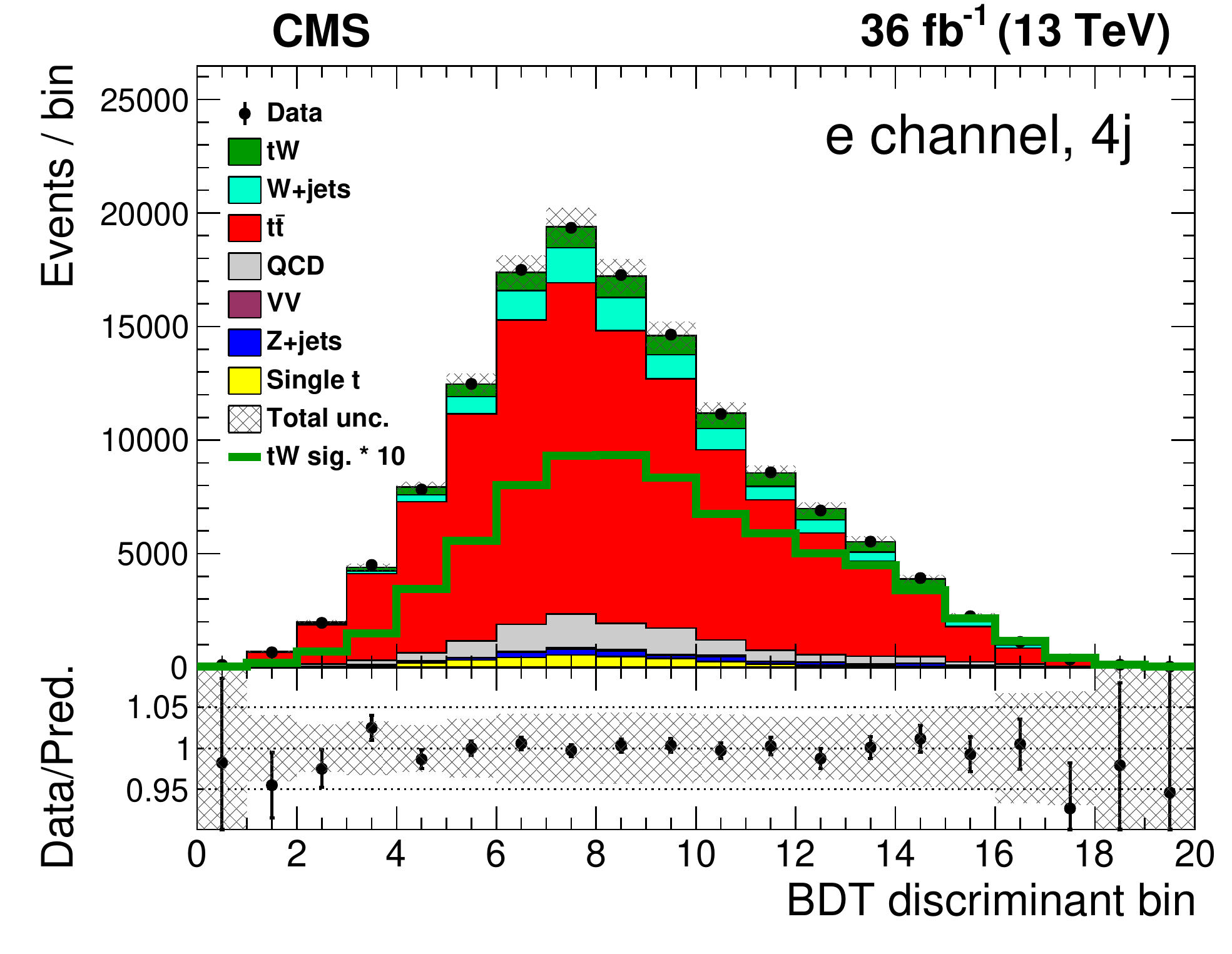}
 \caption{\label{fig:postfitBDT}BDT discriminant in the signal region for the muon (left) and electron (right) channels for the (from upper to lower) 3j, 2j and 4j regions.
     The upper 3j region is considered the nominal signal region, while the remaining 2j and 4j regions are considered control regions, enhanced in \wPlusJets and QCD multijet, and \ttbar background events, respectively.
     The shape of the discriminant for the \tW signal multiplied by 10 is overlayed.
     The signal and backgrounds have been scaled with the results of the fit.
     The lower panel shows the ratio of observed data to the prediction for signal and background.
     In both panels the hatched regions show the total uncertainty of the prediction.
   }
\end{figure}

Figure~\ref{fig:postfitBDT} shows the BDT discriminant for the signal and control regions scaled to the output of the fit.

\section{Summary}

The first observation of the associated production of a single top quark and a \PW boson in the single-lepton channel containing a muon or electron and jets is presented.
The cross section is extracted using a binned likelihood fit of the discriminant from a boosted decision tree designed to separate the signal from the dominant top quark and antiquark pair background.
The analysis is performed using proton-proton collision data at a centre-of-mass energy of 13\TeV recorded by the CMS detector at the LHC corresponding to an integrated luminosity of \Lumimine.

The cross section is \xsecmes, with a significance exceeding 5 standard deviations, which is compatible with both the standard model predictions at approximate next-to-next-to-leading order in quantum chromodynamics of \xsecannlo and at approximate next-to-next-to-next-to-leading order of \xsecannnlo.

\begin{acknowledgments}
      We congratulate our colleagues in the CERN accelerator departments for the excellent performance of the LHC and thank the technical and administrative staffs at CERN and at other CMS institutes for their contributions to the success of the CMS effort. In addition, we gratefully acknowledge the computing centres and personnel of the Worldwide LHC Computing Grid and other centres for delivering so effectively the computing infrastructure essential to our analyses. Finally, we acknowledge the enduring support for the construction and operation of the LHC, the CMS detector, and the supporting computing infrastructure provided by the following funding agencies: BMBWF and FWF (Austria); FNRS and FWO (Belgium); CNPq, CAPES, FAPERJ, FAPERGS, and FAPESP (Brazil); MES and BNSF (Bulgaria); CERN; CAS, MoST, and NSFC (China); MINCIENCIAS (Colombia); MSES and CSF (Croatia); RIF (Cyprus); SENESCYT (Ecuador); MoER, ERC PUT and ERDF (Estonia); Academy of Finland, MEC, and HIP (Finland); CEA and CNRS/IN2P3 (France); BMBF, DFG, and HGF (Germany); GSRI (Greece); NKFIA (Hungary); DAE and DST (India); IPM (Iran); SFI (Ireland); INFN (Italy); MSIP and NRF (Republic of Korea); MES (Latvia); LAS (Lithuania); MOE and UM (Malaysia); BUAP, CINVESTAV, CONACYT, LNS, SEP, and UASLP-FAI (Mexico); MOS (Montenegro); MBIE (New Zealand); PAEC (Pakistan); MSHE and NSC (Poland); FCT (Portugal); JINR (Dubna); MON, RosAtom, RAS, RFBR, and NRC KI (Russia); MESTD (Serbia); SEIDI, CPAN, PCTI, and FEDER (Spain); MOSTR (Sri Lanka); Swiss Funding Agencies (Switzerland); MST (Taipei); ThEPCenter, IPST, STAR, and NSTDA (Thailand); TUBITAK and TAEK (Turkey); NASU (Ukraine); STFC (United Kingdom); DOE and NSF (USA).
       
      \hyphenation{Rachada-pisek} Individuals have received support from the Marie-Curie programme and the European Research Council and Horizon 2020 Grant, contract Nos.\ 675440, 724704, 752730, 758316, 765710, 824093, 884104, and COST Action CA16108 (European Union); the Leventis Foundation; the Alfred P.\ Sloan Foundation; the Alexander von Humboldt Foundation; the Belgian Federal Science Policy Office; the Fonds pour la Formation \`a la Recherche dans l'Industrie et dans l'Agriculture (FRIA-Belgium); the Agentschap voor Innovatie door Wetenschap en Technologie (IWT-Belgium); the F.R.S.-FNRS and FWO (Belgium) under the ``Excellence of Science -- EOS" -- be.h project n.\ 30820817; the Beijing Municipal Science \& Technology Commission, No. Z191100007219010; the Ministry of Education, Youth and Sports (MEYS) of the Czech Republic; the Deutsche Forschungsgemeinschaft (DFG), under Germany's Excellence Strategy -- EXC 2121 ``Quantum Universe" -- 390833306, and under project number 400140256 - GRK2497; the Lend\"ulet (``Momentum") Programme and the J\'anos Bolyai Research Scholarship of the Hungarian Academy of Sciences, the New National Excellence Program \'UNKP, the NKFIA research grants 123842, 123959, 124845, 124850, 125105, 128713, 128786, and 129058 (Hungary); the Council of Science and Industrial Research, India; the Latvian Council of Science; the Ministry of Science and Higher Education and the National Science Center, contracts Opus 2014/15/B/ST2/03998 and 2015/19/B/ST2/02861 (Poland); the Funda\c{c}\~ao para a Ci\^encia e a Tecnologia, grant CEECIND/01334/2018 (Portugal); the National Priorities Research Program by Qatar National Research Fund; the Ministry of Science and Higher Education, project no. 14.W03.31.0026 (Russia); the Programa Estatal de Fomento de la Investigaci{\'o}n Cient{\'i}fica y T{\'e}cnica de Excelencia Mar\'{\i}a de Maeztu, grant MDM-2015-0509 and the Programa Severo Ochoa del Principado de Asturias; the Stavros Niarchos Foundation (Greece); the Rachadapisek Sompot Fund for Postdoctoral Fellowship, Chulalongkorn University and the Chulalongkorn Academic into Its 2nd Century Project Advancement Project (Thailand); the Kavli Foundation; the Nvidia Corporation; the SuperMicro Corporation; the Welch Foundation, contract C-1845; and the Weston Havens Foundation (USA). 
\end{acknowledgments}

\bibliography{auto_generated}   

\providecommand{\href}[2]{#2}\begingroup\raggedright\begin{thebibliography}{10}%
\makeatletter
\providecommand{\hrefCMSnoop }[0]{\@secondoftwo}%
\makeatother
\providecommand{\doi}{\texttt{doi:}\begingroup \urlstyle{tt}\Url}

\bibitem{PhysRevLett.103.092001}
\hrefCMSnoop {}{{D0} Collaboration, ``Observation of single top-quark
  production'',} \textit{ Phys. Rev. Lett.} \textbf{ 103} (2009) 092001,
  \href{http://dx.doi.org/10.1103/PhysRevLett.103.092001}{\doi{10.1103/PhysRevLett.103.092001}},
  \href{http://www.arXiv.org/abs/0903.0850}{\texttt{arXiv:0903.0850}}.

\bibitem{PhysRevLett.103.092002}
\hrefCMSnoop {}{{CDF} Collaboration, ``Observation of electroweak single
  top-quark production'',} \textit{ Phys. Rev. Lett.} \textbf{ 103} (2009)
  092002,
  \href{http://dx.doi.org/10.1103/PhysRevLett.103.092002}{\doi{10.1103/PhysRevLett.103.092002}},
  \href{http://www.arXiv.org/abs/0903.0885}{\texttt{arXiv:0903.0885}}.

\bibitem{ATLAS:2016qhd}
\hrefCMSnoop {}{{ATLAS Collaboration}, ``Measurement of the inclusive
  cross-sections of single top-quark and top-antiquark $t$-channel production
  in $pp$ collisions at $\sqrt{s} = 13$ {TeV} with the {ATLAS} detector'',}
  \textit{ JHEP} \textbf{ 04} (2017) 086,
  \href{http://dx.doi.org/10.1007/JHEP04(2017)086}{\doi{10.1007/JHEP04(2017)086}},
  \href{http://www.arXiv.org/abs/1609.03920}{\texttt{arXiv:1609.03920}}.

\bibitem{CMS:2018lgn}
\hrefCMSnoop {}{{CMS Collaboration}, ``Measurement of the single top quark and
  antiquark production cross sections in the $t$ channel and their ratio in
  proton-proton collisions at $\sqrt{s}= 13$ {TeV}'',} \textit{ Phys. Lett. B}
  \textbf{ 800} (2020) 135042,
  \href{http://dx.doi.org/10.1016/j.physletb.2019.135042}{\doi{10.1016/j.physletb.2019.135042}},
  \href{http://www.arXiv.org/abs/1812.10514}{\texttt{arXiv:1812.10514}}.

\bibitem{ATLAS:2019hhu}
\hrefCMSnoop {}{{ATLAS, CMS} Collaboration, ``Combinations of single-top-quark
  production cross-section measurements and $\abs{f_{LV}V_{tb}}$ determinations
  at $ \sqrt{s} = 7$ and 8 {TeV} with the {ATLAS} and {CMS} experiments'',}
  \textit{ JHEP} \textbf{ 05} (2019) 088,
  \href{http://dx.doi.org/10.1007/JHEP05(2019)088}{\doi{10.1007/JHEP05(2019)088}},
  \href{http://www.arXiv.org/abs/1902.07158}{\texttt{arXiv:1902.07158}}.

\bibitem{ATLAS:2015jmq}
\hrefCMSnoop {}{{ATLAS Collaboration}, ``Evidence for single top-quark
  production in the $s$-channel in proton-proton collisions at $\sqrt{s}= 8$
  {TeV} with the {ATLAS} detector using the {Matrix Element Method}'',}
  \textit{ Phys. Lett. B} \textbf{ 756} (2016) 228,
  \href{http://dx.doi.org/10.1016/j.physletb.2016.03.017}{\doi{10.1016/j.physletb.2016.03.017}},
  \href{http://www.arXiv.org/abs/1511.05980}{\texttt{arXiv:1511.05980}}.

\bibitem{CMS:2016xoq}
\hrefCMSnoop {}{{CMS Collaboration}, ``Search for s channel single top quark
  production in pp collisions at $ \sqrt{s}=7 $ and 8 {TeV}'',} \textit{ JHEP}
  \textbf{ 09} (2016) 027,
  \href{http://dx.doi.org/10.1007/JHEP09(2016)027}{\doi{10.1007/JHEP09(2016)027}},
  \href{http://www.arXiv.org/abs/1603.02555}{\texttt{arXiv:1603.02555}}.

\bibitem{Tait:2000sh}
\hrefCMSnoop {}{T.~M.~P. Tait and C.~P. Yuan, ``Single top quark production as
  a window to physics beyond the standard model'',} \textit{ Phys. Rev. D}
  \textbf{ 63} (2000) 014018,
  \href{http://dx.doi.org/10.1103/PhysRevD.63.014018}{\doi{10.1103/PhysRevD.63.014018}},
\href{http://www.arXiv.org/abs/hep-ph/0007298}{\texttt{arXiv:hep-ph/0007298}}.
%%CITATION = HEP-PH/0007298;%%.

\bibitem{Cao:2007ea}
\hrefCMSnoop {}{Q.-H. Cao, J.~Wudka, and C.~P. Yuan, ``Search for new physics
  via single top production at the {LHC}'',} \textit{ Phys. Lett. B} \textbf{
  658} (2007) 50,
  \href{http://dx.doi.org/10.1016/j.physletb.2007.10.057}{\doi{10.1016/j.physletb.2007.10.057}},
\href{http://www.arXiv.org/abs/0704.2809}{\texttt{arXiv:0704.2809}}.
%%CITATION = ARXIV:0704.2809;%%.

\bibitem{Aad:2012xca}
\hrefCMSnoop {}{{ATLAS Collaboration}, ``Evidence for the associated production
  of a {\PW} boson and a top quark in {ATLAS} at $\sqrt{s}=7$ {TeV}'',}
  \textit{ Phys. Lett. B} \textbf{ 716} (2012) 142,
  \href{http://dx.doi.org/10.1016/j.physletb.2012.08.011}{\doi{10.1016/j.physletb.2012.08.011}},
  \href{http://www.arXiv.org/abs/1205.5764}{\texttt{arXiv:1205.5764}}.

\bibitem{Chatrchyan:2012zca}
\hrefCMSnoop {}{{CMS Collaboration}, ``Evidence for associated production of a
  single top quark and {\PW} boson in \pp collisions at $\sqrt{s} = 7$
  {TeV}'',} \textit{ Phys. Rev. Lett.} \textbf{ 110} (2013) 022003,
  \href{http://dx.doi.org/10.1103/PhysRevLett.110.022003}{\doi{10.1103/PhysRevLett.110.022003}},
  \href{http://www.arXiv.org/abs/1209.3489}{\texttt{arXiv:1209.3489}}.

\bibitem{Chatrchyan:2014tua}
\hrefCMSnoop {}{{CMS Collaboration}, ``Observation of the associated production
  of a single top quark and a {\PW} boson in \pp collisions at $\sqrt{s} = 8$
  {TeV}'',} \textit{ Phys. Rev. Lett.} \textbf{ 112} (2014) 231802,
  \href{http://dx.doi.org/10.1103/PhysRevLett.112.231802}{\doi{10.1103/PhysRevLett.112.231802}},
\href{http://www.arXiv.org/abs/1401.2942}{\texttt{arXiv:1401.2942}}.
%%CITATION = ARXIV:1401.2942;%%.

\bibitem{Aad:2015eto}
\hrefCMSnoop {}{{ATLAS Collaboration}, ``Measurement of the production
  cross-section of a single top quark in association with a {\PW} boson at 8
  {TeV} with the {ATLAS} experiment'',} \textit{ JHEP} \textbf{ 01} (2016) 064,
  \href{http://dx.doi.org/10.1007/JHEP01(2016)064}{\doi{10.1007/JHEP01(2016)064}},
  \href{http://www.arXiv.org/abs/1510.03752}{\texttt{arXiv:1510.03752}}.

\bibitem{Aaboud:2016lpj}
\hrefCMSnoop {}{{ATLAS Collaboration}, ``Measurement of the cross-section for
  producing a {\PW} boson in association with a single top quark in \pp
  collisions at $ \sqrt{s}=13 $ {TeV} with {ATLAS}'',} \textit{ JHEP} \textbf{
  01} (2018) 063,
  \href{http://dx.doi.org/10.1007/JHEP01(2018)063}{\doi{10.1007/JHEP01(2018)063}},
  \href{http://www.arXiv.org/abs/1612.07231}{\texttt{arXiv:1612.07231}}.

\bibitem{Aaboud:2017qyi}
\hrefCMSnoop {}{{ATLAS Collaboration}, ``Measurement of differential
  cross-sections of a single top quark produced in association with a {\PW}
  boson at $\sqrt{s}=13$ {TeV} with {ATLAS}'',} \textit{ Eur. Phys. J. C}
  \textbf{ 78} (2018) 186,
  \href{http://dx.doi.org/10.1140/epjc/s10052-018-5649-8}{\doi{10.1140/epjc/s10052-018-5649-8}},
  \href{http://www.arXiv.org/abs/1712.01602}{\texttt{arXiv:1712.01602}}.

\bibitem{Sirunyan:2018lcp}
\hrefCMSnoop {}{{CMS Collaboration}, ``Measurement of the production cross
  section for single top quarks in association with {\PW} bosons in
  proton-proton collisions at $ \sqrt{s}=13 $ {TeV}'',} \textit{ JHEP} \textbf{
  10} (2018) 117,
  \href{http://dx.doi.org/10.1007/JHEP10(2018)117}{\doi{10.1007/JHEP10(2018)117}},
  \href{http://www.arXiv.org/abs/1805.07399}{\texttt{arXiv:1805.07399}}.

\bibitem{Kidonakis:2015nna}
\hrefCMSnoop {}{N.~Kidonakis, ``Theoretical results for electroweak-boson and
  single-top production'',} \textit{ PoS} \textbf{ DIS2015} (2015) 170,
  \href{http://dx.doi.org/10.22323/1.247.0170}{\doi{10.22323/1.247.0170}},
  \href{http://www.arXiv.org/abs/1506.04072}{\texttt{arXiv:1506.04072}}.

\bibitem{Kidonakis_2021}
\hrefCMSnoop {}{N.~Kidonakis and N.~Yamanaka, ``Higher-order corrections for
  {\tW} production at high-energy hadron colliders'',} \textit{ JHEP} \textbf{
  05} (2021) 278,
  \href{http://dx.doi.org/10.1007/jhep05(2021)278}{\doi{10.1007/jhep05(2021)278}},
  \href{http://www.arXiv.org/abs/2102.11300}{\texttt{arXiv:2102.11300}}.

\bibitem{PhysRevD.59.075001}
\hrefCMSnoop {}{A.~S. Belyaev, E.~E. Boos, and L.~V. Dudko, ``Single top quark
  at future hadron colliders: Complete signal and background study'',} \textit{
  Phys. Rev. D} \textbf{ 59} (1999) 075001,
  \href{http://dx.doi.org/10.1103/PhysRevD.59.075001}{\doi{10.1103/PhysRevD.59.075001}},
  \href{http://www.arXiv.org/abs/hep-ph/9806332}{\texttt{arXiv:hep-ph/9806332}}.

\bibitem{White:2009yt}
\hrefCMSnoop {}{C.~D. White, S.~Frixione, E.~Laenen, and F.~Maltoni,
  ``Isolating {Wt} production at the {LHC}'',} \textit{ JHEP} \textbf{ 11}
  (2009) 074,
  \href{http://dx.doi.org/10.1088/1126-6708/2009/11/074}{\doi{10.1088/1126-6708/2009/11/074}},
  \href{http://www.arXiv.org/abs/0908.0631}{\texttt{arXiv:0908.0631}}.

\bibitem{Frixione:2008yi}
S.~Frixione\hrefCMSnoop {}{ {et~al.}, ``Single-top hadroproduction in
  association with a {\PW} boson'',} \textit{ JHEP} \textbf{ 07} (2008) 029,
  \href{http://dx.doi.org/10.1088/1126-6708/2008/07/029}{\doi{10.1088/1126-6708/2008/07/029}},
  \href{http://www.arXiv.org/abs/0805.3067}{\texttt{arXiv:0805.3067}}.

\bibitem{Tait:1999cf}
\hrefCMSnoop {}{T.~M.~P. Tait, ``{$\PQt\PW^{-}$} mode of single top
  production'',} \textit{ Phys. Rev. D} \textbf{ 61} (1999) 034001,
  \href{http://dx.doi.org/10.1103/PhysRevD.61.034001}{\doi{10.1103/PhysRevD.61.034001}},
\href{http://www.arXiv.org/abs/hep-ph/9909352}{\texttt{arXiv:hep-ph/9909352}}.
%%CITATION = HEP-PH/9909352;%%.

\bibitem{atlascollaboration2020measurement}
\hrefCMSnoop {}{{ATLAS Collaboration}, ``Measurement of single top-quark
  production in association with a {\PW} boson in the single-lepton channel at
  $\sqrt{s} = 8$ {TeV} with the {ATLAS} detector'',} \textit{ Eur. Phys. J. C}
  \textbf{ 81} (2021) 720,
  \href{http://dx.doi.org/10.1140/epjc/s10052-021-09371-7}{\doi{10.1140/epjc/s10052-021-09371-7}},
  \href{http://www.arXiv.org/abs/2007.01554}{\texttt{arXiv:2007.01554}}.

\bibitem{hepDataEntry}
\hrefCMSnoop {}{``{HEPData} record for this analysis'',} 2021.
\newblock
  \href{http://dx.doi.org/10.17182/hepdata.102957}{\doi{10.17182/hepdata.102957}}.

\bibitem{Cacciari:2008gp}
\hrefCMSnoop {}{M.~Cacciari, G.~P. Salam, and G.~Soyez, ``The anti-\kt jet
  clustering algorithm'',} \textit{ JHEP} \textbf{ 04} (2008) 063,
  \href{http://dx.doi.org/10.1088/1126-6708/2008/04/063}{\doi{10.1088/1126-6708/2008/04/063}},
  \href{http://www.arXiv.org/abs/0802.1189}{\texttt{arXiv:0802.1189}}.

\bibitem{Cacciari:2011ma}
\hrefCMSnoop {}{M.~Cacciari, G.~P. Salam, and G.~Soyez, ``{FastJet} user
  manual'',} \textit{ Eur. Phys. J. C} \textbf{ 72} (2012) 1896,
  \href{http://dx.doi.org/10.1140/epjc/s10052-012-1896-2}{\doi{10.1140/epjc/s10052-012-1896-2}},
\href{http://www.arXiv.org/abs/1111.6097}{\texttt{arXiv:1111.6097}}.
%%CITATION = ARXIV:1111.6097;%%.

\bibitem{CMS-PRF-14-001}
\hrefCMSnoop {}{{CMS Collaboration}, ``Particle-flow reconstruction and global
  event description with the {CMS} detector'',} \textit{ JINST} \textbf{ 12}
  (2017) P10003,
  \href{http://dx.doi.org/10.1088/1748-0221/12/10/P10003}{\doi{10.1088/1748-0221/12/10/P10003}},
\href{http://www.arXiv.org/abs/1706.04965}{\texttt{arXiv:1706.04965}}.
%%CITATION = ARXIV:1706.04965;%%.

\bibitem{muonRecoPaper}
\hrefCMSnoop {}{{CMS Collaboration}, ``Performance of the {CMS} muon detector
  and muon reconstruction with proton-proton collisions at $\sqrt{s}=13$
  {TeV}'',} \textit{ JINST} \textbf{ 13} (2018) P06015,
  \href{http://dx.doi.org/10.1088/1748-0221/13/06/p06015}{\doi{10.1088/1748-0221/13/06/p06015}},
  \href{http://www.arXiv.org/abs/1804.04528}{\texttt{arXiv:1804.04528}}.

\bibitem{Sirunyan:2019kia}
\hrefCMSnoop {}{{CMS Collaboration}, ``Performance of missing transverse
  momentum reconstruction in proton-proton collisions at $\sqrt{s} = 13$ {TeV}
  using the {CMS} detector'',} \textit{ JINST} \textbf{ 14} (2019) P07004,
  \href{http://dx.doi.org/10.1088/1748-0221/14/07/P07004}{\doi{10.1088/1748-0221/14/07/P07004}},
\href{http://www.arXiv.org/abs/1903.06078}{\texttt{arXiv:1903.06078}}.
%%CITATION = ARXIV:1903.06078;%%.

\bibitem{Chatrchyan:2008zzk}
\hrefCMSnoop {}{{CMS Collaboration}, ``The {CMS} experiment at the {CERN}
  {LHC}'',} \textit{ JINST} \textbf{ 3} (2008) S08004,
  \href{http://dx.doi.org/10.1088/1748-0221/3/08/S08004}{\doi{10.1088/1748-0221/3/08/S08004}}.

\bibitem{lumiPaper}
\href {http://cds.cern.ch/record/2257069}{{{CMS}} Collaboration, ``{CMS}
  luminosity measurements for the 2016 data taking period'',} CMS Physics
  Analysis Summary CMS-PAS-LUM-17-001, 2017.

\bibitem{Re:2010bp}
\hrefCMSnoop {}{E.~Re, ``Single-top {Wt}-channel production matched with parton
  showers using the {POWHEG} method'',} \textit{ Eur. Phys. J. C} \textbf{ 71}
  (2011) 1547,
  \href{http://dx.doi.org/10.1140/epjc/s10052-011-1547-z}{\doi{10.1140/epjc/s10052-011-1547-z}},
\href{http://www.arXiv.org/abs/1009.2450}{\texttt{arXiv:1009.2450}}.
%%CITATION = ARXIV:1009.2450;%%.

\bibitem{Sjostrand:2014zea}
T.~Sj{\"o}strand\hrefCMSnoop {}{ {et~al.}, ``An introduction to {PYTHIA}
  8.2'',} \textit{ Comput. Phys. Commun.} \textbf{ 191} (2015) 159,
  \href{http://dx.doi.org/10.1016/j.cpc.2015.01.024}{\doi{10.1016/j.cpc.2015.01.024}},
\href{http://www.arXiv.org/abs/1410.3012}{\texttt{arXiv:1410.3012}}.
%%CITATION = ARXIV:1410.3012;%%.

\bibitem{Khachatryan:2015pea}
\hrefCMSnoop {}{{CMS Collaboration}, ``Event generator tunes obtained from
  underlying event and multiparton scattering measurements'',} \textit{ Eur.
  Phys. J. C} \textbf{ 76} (2016) 155,
  \href{http://dx.doi.org/10.1140/epjc/s10052-016-3988-x}{\doi{10.1140/epjc/s10052-016-3988-x}},
\href{http://www.arXiv.org/abs/1512.00815}{\texttt{arXiv:1512.00815}}.
%%CITATION = ARXIV:1512.00815;%%.

\bibitem{Alioli:2010xd}
\hrefCMSnoop {}{S.~Alioli, P.~Nason, C.~Oleari, and E.~Re, ``A general
  framework for implementing {NLO} calculations in shower {Monte Carlo}
  programs: the {POWHEG BOX}'',} \textit{ JHEP} \textbf{ 06} (2010) 043,
  \href{http://dx.doi.org/10.1007/JHEP06(2010)043}{\doi{10.1007/JHEP06(2010)043}},
\href{http://www.arXiv.org/abs/1002.2581}{\texttt{arXiv:1002.2581}}.
%%CITATION = ARXIV:1002.2581;%%.

\bibitem{Skands:2014pea}
\hrefCMSnoop {}{P.~Skands, S.~Carrazza, and J.~Rojo, ``Tuning {PYTHIA} 8.1: the
  {Monash 2013 Tune}'',} \textit{ Eur. Phys. J. C} \textbf{ 74} (2014) 3024,
  \href{http://dx.doi.org/10.1140/epjc/s10052-014-3024-y}{\doi{10.1140/epjc/s10052-014-3024-y}},
\href{http://www.arXiv.org/abs/1404.5630}{\texttt{arXiv:1404.5630}}.
%%CITATION = ARXIV:1404.5630;%%.

\bibitem{Alwall:2014hca}
J.~Alwall\hrefCMSnoop {}{ {et~al.}, ``The automated computation of tree-level
  and next-to-leading order differential cross sections, and their matching to
  parton shower simulations'',} \textit{ JHEP} \textbf{ 07} (2014) 079,
  \href{http://dx.doi.org/10.1007/JHEP07(2014)079}{\doi{10.1007/JHEP07(2014)079}},
\href{http://www.arXiv.org/abs/1405.0301}{\texttt{arXiv:1405.0301}}.
%%CITATION = ARXIV:1405.0301;%%.

\bibitem{Frederix:2012ps}
\hrefCMSnoop {}{R.~Frederix and S.~Frixione, ``Merging meets matching in
  {MC@NLO}'',} \textit{ JHEP} \textbf{ 12} (2012) 061,
  \href{http://dx.doi.org/10.1007/JHEP12(2012)061}{\doi{10.1007/JHEP12(2012)061}},
\href{http://www.arXiv.org/abs/1209.6215}{\texttt{arXiv:1209.6215}}.
%%CITATION = ARXIV:1209.6215;%%.

\bibitem{Artoisenet:2012st}
\hrefCMSnoop {}{P.~Artoisenet, R.~Frederix, O.~Mattelaer, and R.~Rietkerk,
  ``Automatic spin-entangled decays of heavy resonances in {Monte Carlo}
  simulations'',} \textit{ JHEP} \textbf{ 03} (2013) 015,
  \href{http://dx.doi.org/10.1007/JHEP03(2013)015}{\doi{10.1007/JHEP03(2013)015}},
\href{http://www.arXiv.org/abs/1212.3460}{\texttt{arXiv:1212.3460}}.
%%CITATION = ARXIV:1212.3460;%%.

\bibitem{Alwall:2007fs}
J.~Alwall\hrefCMSnoop {}{ {et~al.}, ``Comparative study of various algorithms
  for the merging of parton showers and matrix elements in hadronic
  collisions'',} \textit{ Eur. Phys. J. C} \textbf{ 53} (2008) 473,
  \href{http://dx.doi.org/10.1140/epjc/s10052-007-0490-5}{\doi{10.1140/epjc/s10052-007-0490-5}},
\href{http://www.arXiv.org/abs/0706.2569}{\texttt{arXiv:0706.2569}}.
%%CITATION = ARXIV:0706.2569;%%.

\bibitem{Ball:2012cx}
\hrefCMSnoop {}{{NNPDF} Collaboration, ``Parton distributions with {LHC}
  data'',} \textit{ Nucl. Phys. B} \textbf{ 867} (2013) 244,
  \href{http://dx.doi.org/10.1016/j.nuclphysb.2012.10.003}{\doi{10.1016/j.nuclphysb.2012.10.003}},
\href{http://www.arXiv.org/abs/1207.1303}{\texttt{arXiv:1207.1303}}.
%%CITATION = ARXIV:1207.1303;%%.

\bibitem{AGOSTINELLI2003250}
\hrefCMSnoop {}{{GEANT4} Collaboration, ``{\GEANTfour}---a simulation
  toolkit'',} \textit{ Nucl. Instrum. Meth. A} \textbf{ 506} (2003) 250,
\href{http://dx.doi.org/10.1016/S0168-9002(03)01368-8}{\doi{10.1016/S0168-9002(03)01368-8}}.
%%CITATION = NUIMA,A506,250;%%.

\bibitem{1610988}
\hrefCMSnoop {}{J.~Allison {et~al.}, ``{\GEANTfour} developments and
  applications'',} \textit{ IEEE Trans. Nucl. Sci.} \textbf{ 53} (2006) 270,
  \href{http://dx.doi.org/10.1109/TNS.2006.869826}{\doi{10.1109/TNS.2006.869826}}.

\bibitem{Sirunyan:2020zal}
\hrefCMSnoop {}{{CMS Collaboration}, ``Performance of the {CMS Level-1} trigger
  in proton-proton collisions at $\sqrt{s} = 13$ {TeV}'',} \textit{ JINST}
  \textbf{ 15} (2020) P10017,
  \href{http://dx.doi.org/10.1088/1748-0221/15/10/P10017}{\doi{10.1088/1748-0221/15/10/P10017}},
  \href{http://www.arXiv.org/abs/2006.10165}{\texttt{arXiv:2006.10165}}.

\bibitem{Khachatryan:2016bia}
\hrefCMSnoop {}{{CMS Collaboration}, ``The {CMS} trigger system'',} \textit{
  JINST} \textbf{ 12} (2017) P01020,
  \href{http://dx.doi.org/10.1088/1748-0221/12/01/P01020}{\doi{10.1088/1748-0221/12/01/P01020}},
\href{http://www.arXiv.org/abs/1609.02366}{\texttt{arXiv:1609.02366}}.
%%CITATION = ARXIV:1609.02366;%%.

\bibitem{Khachatryan:2015hwa}
\hrefCMSnoop {}{{CMS Collaboration}, ``Performance of electron reconstruction
  and selection with the {CMS} detector in proton-proton collisions at
  $\sqrt{s} = 8$ {TeV}'',} \textit{ JINST} \textbf{ 10} (2015) P06005,
  \href{http://dx.doi.org/10.1088/1748-0221/10/06/P06005}{\doi{10.1088/1748-0221/10/06/P06005}},
\href{http://www.arXiv.org/abs/1502.02701}{\texttt{arXiv:1502.02701}}.
%%CITATION = ARXIV:1502.02701;%%.

\bibitem{Khachatryan:2016kdb}
\hrefCMSnoop {}{{CMS Collaboration}, ``Jet energy scale and resolution in the
  {CMS} experiment in \pp collisions at 8 tev'',} \textit{ JINST} \textbf{ 12}
  (2017) P02014,
  \href{http://dx.doi.org/10.1088/1748-0221/12/02/P02014}{\doi{10.1088/1748-0221/12/02/P02014}},
\href{http://www.arXiv.org/abs/1607.03663}{\texttt{arXiv:1607.03663}}.
%%CITATION = ARXIV:1607.03663;%%.

\bibitem{Sirunyan:2018bby}
\hrefCMSnoop {}{{CMS Collaboration}, ``Identification of heavy-flavour jets
  with the {CMS} detector in \pp collisions at 13 {TeV}'',} \textit{ JINST}
  \textbf{ 13} (2018) P05011,
  \href{http://dx.doi.org/10.1088/1748-0221/13/05/p05011}{\doi{10.1088/1748-0221/13/05/p05011}},
  \href{http://www.arXiv.org/abs/1712.07158}{\texttt{arXiv:1712.07158}}.

\bibitem{Hocker:2007ht}
\hrefCMSnoop {}{H.~Voss, A.~H{\"o}cker, J.~Stelzer, and F.~Tegenfeldt,
  ``{TMVA}, the toolkit for multivariate data analysis with {ROOT}'',} in
  \textit{ XIth International Workshop on Advanced Computing and Analysis
  Techniques in Physics Research (ACAT)}, p.~40.
\newblock 2007.
\newblock
  \href{http://www.arXiv.org/abs/physics/0703039}{\texttt{arXiv:physics/0703039}}.
\newblock {[PoS(ACAT)040]}.
\href{http://dx.doi.org/10.22323/1.050.0040}{\doi{10.22323/1.050.0040}}.
%%CITATION = PHYSICS/0703039;%%.

\bibitem{Yang:2005nz}
\hrefCMSnoop {}{H.-J. Yang, B.~P. Roe, and J.~Zhu, ``Studies of boosted
  decision trees for {MiniBooNE} particle identification'',} \textit{ Nucl.
  Instrum. Meth. A} \textbf{ 555} (2005) 370,
  \href{http://dx.doi.org/10.1016/j.nima.2005.09.022}{\doi{10.1016/j.nima.2005.09.022}},
  \href{http://www.arXiv.org/abs/physics/0508045}{\texttt{arXiv:physics/0508045}}.

\bibitem{Sirunyan:2018nqx}
\hrefCMSnoop {}{{CMS Collaboration}, ``Measurement of the inelastic
  proton-proton cross section at $ \sqrt{s}=13 $ {TeV}'',} \textit{ JHEP}
  \textbf{ 07} (2018) 161,
  \href{http://dx.doi.org/10.1007/JHEP07(2018)161}{\doi{10.1007/JHEP07(2018)161}},
\href{http://www.arXiv.org/abs/1802.02613}{\texttt{arXiv:1802.02613}}.
%%CITATION = ARXIV:1802.02613;%%.

\bibitem{CMS-PAS-JME-16-003}
\href {http://cdsweb.cern.ch/record/2256875}{{{CMS}} Collaboration, ``Jet
  algorithms performance in 13 {TeV} data'',} CMS Physics Analysis Summary
  CMS-PAS-JME-16-003, 2017.

\bibitem{ttbarCrossSection}
\hrefCMSnoop {}{{CMS Collaboration}, ``Measurement of the \ttbar production
  cross section, the top quark mass, and the strong coupling constant using
  dilepton events in \pp collisions at $\sqrt{s}=13$ {TeV}'',} \textit{ Eur.
  Phys. J. C} \textbf{ 79} (2019) 368,
  \href{http://dx.doi.org/10.1140/epjc/s10052-019-6863-8}{\doi{10.1140/epjc/s10052-019-6863-8}},
  \href{http://www.arXiv.org/abs/1812.10505}{\texttt{arXiv:1812.10505}}.

\bibitem{tChanCrossSection}
\hrefCMSnoop {}{{CMS Collaboration}, ``Cross section measurement of $t$-channel
  single top quark production in \pp collisions at $\sqrt{s}=13$ {TeV}'',}
  \textit{ Phys. Lett. B} \textbf{ 772} (2017) 752,
  \href{http://dx.doi.org/10.1016/j.physletb.2017.07.047}{\doi{10.1016/j.physletb.2017.07.047}},
  \href{http://www.arXiv.org/abs/1711.01769}{\texttt{arXiv:1711.01769}}.

\bibitem{Catani:2003zt}
\hrefCMSnoop {}{S.~Catani, D.~de~Florian, M.~Grazzini, and P.~Nason,
  ``Soft-gluon resummation for {Higgs} boson production at hadron colliders'',}
  \textit{ JHEP} \textbf{ 07} (2003) 028,
  \href{http://dx.doi.org/10.1088/1126-6708/2003/07/028}{\doi{10.1088/1126-6708/2003/07/028}},
\href{http://www.arXiv.org/abs/hep-ph/0306211}{\texttt{arXiv:hep-ph/0306211}}.
%%CITATION = HEP-PH/0306211;%%.

\bibitem{Sirunyan:2019dfx}
\hrefCMSnoop {}{{CMS Collaboration}, ``Extraction and validation of a new set
  of {CMS PYTHIA8} tunes from underlying-event measurements'',} \textit{ Eur.
  Phys. J. C} \textbf{ 80} (2020) 4,
  \href{http://dx.doi.org/10.1140/epjc/s10052-019-7499-4}{\doi{10.1140/epjc/s10052-019-7499-4}},
\href{http://www.arXiv.org/abs/1903.12179}{\texttt{arXiv:1903.12179}}.
%%CITATION = ARXIV:1903.12179;%%.

\bibitem{Argyropoulos:2014zoa}
\hrefCMSnoop {}{S.~Argyropoulos and T.~Sj{\"o}strand, ``Effects of color
  reconnection on \ttbar final states at the {LHC}'',} \textit{ JHEP} \textbf{
  11} (2014) 043,
  \href{http://dx.doi.org/10.1007/JHEP11(2014)043}{\doi{10.1007/JHEP11(2014)043}},
\href{http://www.arXiv.org/abs/1407.6653}{\texttt{arXiv:1407.6653}}.
%%CITATION = ARXIV:1407.6653;%%.

\bibitem{Christiansen:2015yqa}
\hrefCMSnoop {}{J.~R. Christiansen and P.~Z. Skands, ``String formation beyond
  leading colour'',} \textit{ JHEP} \textbf{ 08} (2015) 003,
  \href{http://dx.doi.org/10.1007/JHEP08(2015)003}{\doi{10.1007/JHEP08(2015)003}},
\href{http://www.arXiv.org/abs/1505.01681}{\texttt{arXiv:1505.01681}}.
%%CITATION = ARXIV:1505.01681;%%.

\bibitem{Ball:2014uwa}
\hrefCMSnoop {}{{NNPDF} Collaboration, ``Parton distributions for the {LHC Run
  II}'',} \textit{ JHEP} \textbf{ 04} (2015) 040,
  \href{http://dx.doi.org/10.1007/JHEP04(2015)040}{\doi{10.1007/JHEP04(2015)040}},
\href{http://www.arXiv.org/abs/1410.8849}{\texttt{arXiv:1410.8849}}.
%%CITATION = ARXIV:1410.8849;%%.

\end{thebibliography}\endgroup
\cleardoublepage \appendix\section{The CMS Collaboration \label{app:collab}}\begin{sloppypar}\hyphenpenalty=5000\widowpenalty=500\clubpenalty=5000\vskip\cmsinstskip
\textbf{Yerevan Physics Institute, Yerevan, Armenia}\\*[0pt]
A.~Tumasyan
\vskip\cmsinstskip
\textbf{Institut f\"{u}r Hochenergiephysik, Vienna, Austria}\\*[0pt]
W.~Adam, J.W.~Andrejkovic, T.~Bergauer, S.~Chatterjee, M.~Dragicevic, A.~Escalante~Del~Valle, R.~Fr\"{u}hwirth\cmsAuthorMark{1}, M.~Jeitler\cmsAuthorMark{1}, N.~Krammer, L.~Lechner, D.~Liko, I.~Mikulec, P.~Paulitsch, F.M.~Pitters, J.~Schieck\cmsAuthorMark{1}, R.~Sch\"{o}fbeck, M.~Spanring, S.~Templ, W.~Waltenberger, C.-E.~Wulz\cmsAuthorMark{1}
% ...
\vskip\cmsinstskip
% ...
1:  Also at TU Wien, Wien, Austria\\
% ...
\end{sloppypar}
%%% END EDITABLE REGION %%%
% skeleton_end
\end{document}